\documentclass[journal]{IEEEtran}

\usepackage[dvips]{color}
\usepackage{graphicx}
\usepackage{amsmath}

\begin{document}
%
\title{Secrecy Outage and Diversity Analysis of Cognitive Radio Systems}

%
%

\markboth{IEEE Journal on Selected Areas in Communications (accepted to appear)}%
{Yulong Zou \MakeLowercase{\textit{et al.}}: Secrecy Outage and Diversity Analysis of Cognitive Radio Systems}

\author{Yulong~Zou,~\IEEEmembership{Senior Member,~IEEE,}
        Xuelong~Li,~\IEEEmembership{Fellow,~IEEE,}
        and Ying-Chang~Liang,~\IEEEmembership{Fellow,~IEEE}

\thanks{Manuscript received December 20, 2013; revised April 25, 2014. This work was partially supported by the National Natural Science Foundation of China (Grant Nos. 61302104, 61271240, and 61125106), the Scientific Research Foundation of Nanjing University of Posts and Telecommunications (Grant No. NY213014), and the Shaanxi Key Innovation Team of Science and Technology (Grant No. 2012KCT-04).}
\thanks{Y. Zou (corresponding author) is with the Key Laboratory of Broadband Wireless Communication and Sensor Network Technology, Nanjing University of Posts and Telecommunications, Nanjing 210003, Jiangsu, P. R. China. (Email: yulong.zou@njupt.edu.cn)}
\thanks{X. Li is with the Center for OPTical IMagery Analysis and Learning (OPTIMAL), State Key Laboratory of Transient Optics and Photonics, Xi'an Institute of Optics and Precision Mechanics, Chinese Academy of Sciences, Xi'an 710119, Shaanxi, P. R. China. (Email: xuelong\_li@opt.ac.cn)}
\thanks{Y.-C. Liang is with the Institute For Infocomm Research, Agency for Science, Technology and Research, Singapore. (Email: ycliang@i2r.a-star.edu.sg)}

}

\maketitle

\begin{abstract}
In this paper, we investigate the physical-layer security of a multi-user multi-eavesdropper cognitive radio system, which is composed of multiple cognitive users (CUs) transmitting to a common cognitive base station (CBS), {while multiple eavesdroppers may collaborate with each other or perform independently in intercepting the CUs-CBS transmissions, which are called the coordinated and uncoordinated eavesdroppers, respectively}. Considering multiple CUs available, we propose the round-robin scheduling as well as the optimal and suboptimal user scheduling schemes for improving the security of CUs-CBS transmissions against eavesdropping attacks. Specifically, the optimal user scheduling is designed by assuming that the channel state information (CSI) of all links from CUs to CBS, to primary user (PU) and to eavesdroppers are available. By contrast, the suboptimal user scheduling only requires the CSI of CUs-CBS links without the PU's and eavesdroppers' CSI. We derive closed-form expressions of the secrecy outage probability of these three scheduling schemes in the presence of {the coordinated and uncoordinated eavesdroppers}. We also carry out the secrecy diversity analysis and show that the round-robin scheduling achieves the diversity order of only one, whereas the optimal and suboptimal scheduling schemes obtain the full secrecy diversity, {no matter whether the eavesdroppers collaborate or not. In addition, numerical secrecy outage results demonstrate that for both the coordinated and uncoordinated eavesdroppers, the optimal user scheduling achieves the best security performance and the round-robin scheduling performs the worst.} Finally, upon increasing the number of CUs, the secrecy outage probabilities of the optimal and suboptimal user scheduling schemes both improve significantly.
\end{abstract}


\IEEEpeerreviewmaketitle

\section{Introduction}

\IEEEPARstart {C}{ognitive} radio is widely recognized as a dynamic spectrum access technique, which enables unlicensed users (also called secondary users or cognitive users) and licensed users (known as primary users) to share the same spectrum but with different priorities, where the primary users (PUs) have a higher priority than the cognitive users (CUs) in accessing the licensed spectrum [1]-[3]. In cognitive radio systems, CUs are typically allowed to detect whether or not the licensed spectrum is being used by PUs through spectrum sensing functionality and then to access the detected unused spectrum (referred to as spectrum hole) [4], [5]. Due to the dynamic nature of cognitive radio, various malicious devices may participate in the spectrum sensing and access, leading legitimate users to be exposed to both internal and external attacks [6]. For example, cognitive radio is supposed to be capable of adapting its operating parameters to any changes of its surrounding radio environment. However, a malicious attacker may intentionally modify the radio environment (e.g., by emitting interference) in which the cognitive radio operates, misleading legitimate CUs and even causing them to malfunction. Therefore, cognitive radio faces many new security challenges from all aspects of the networking architecture, including the spectrum sensing, spectrum access, and spectrum management.

Physical-layer security [7]-[9] is emerging as an effective means to protect the communications confidentiality against eavesdropping attacks by exploiting the physical characteristics (e.g., multipath fading, propagation delay, etc.) of wireless channels. It has been shown that if the wiretap channel (from source to eavesdropper) is inferior to the main channel (from source to destination), the source can reliably and securely transmit to the destination at a positive data rate (see [10] and reference therein). In [7], Wyner introduced the notation of secrecy capacity in a discrete memoryless wiretap channel and showed the secrecy capacity as the difference between the capacities of the main channel and wiretap channel. Later on, Wyner's results were extended to the Gaussian wiretap channel in [8] and wireless fading channels in [9] and [11], where the achievable rate-equivocation region was characterized from an information-theoretic perspective. It is noted that the secrecy capacity of wireless communications is limited and degraded due to the multipath fading effect. To this end, considerable research efforts were devoted to improving the wireless physical-layer security by employing the multiple-input multiple-output (MIMO) [12], artificial noise [13], [14] and beamforming techniques [15]-[17]. In addition, the joint artificial noise and beamforming design was investigated in [18] to enhance the wireless physical-layer security, where the artificial noise covariance and beamforming weights were jointly optimized with a target secrecy rate requirement. It was demonstrated that the joint artificial noise and beamforming approach further improves the wireless secrecy capacity.

As aforementioned, the physical-layer security is examined extensively for conventional non-cognitive wireless networks [9]-[18], but is rarely studied for cognitive radio networks. {The physical-layer security of cognitive transmissions was investigated in [19]-[21] where the achievable secrecy rates of the multiple-input single-output (MISO), MIMO and relay selection were developed for cognitive radio networks. More recently, in [22], we examined the physical-layer security with multi-user scheduling for cognitive radio networks in terms of the ergodic secrecy rate and intercept probability. In this paper, we explore the physical-layer security of a multi-user multi-eavesdropper (MUME) cognitive radio network, where the eavesdroppers may collaborate with each other or perform independently in intercepting the cognitive transmissions. This is different from the existing cognitive radio security works [19]-[22] in the following aspects. On the one hand, we examine the use of multi-user scheduling for improving the physical-layer security of cognitive transmissions, whereas multiple antennas or multiple relays are employed in [19]-[21] with the aid of antenna array design or relay selection. On the other hand, we are focused on the secrecy outage probability analysis of cognitive radio networks in the presence of both the uncoordinated and coordinated eavesdroppers, differing from our previous work [22], where the intercept probability of cognitive transmissions was analyzed for the uncoordinated eavesdroppers only. Notice that the intercept probability was defined in [22] as the probability that the capacity of the main channel falls below that of the wiretap channel. By contrast, the secrecy outage probability is the probability that the difference between the capacity of the main channel and that of the wiretap channel becomes less than a predefined secrecy rate (i.e., $R_s$). It can be observed that the intercept probability is just a special case of the secrecy outage probability with $R_s=0$, showing that the secrecy outage probability to be studied in this paper is more general than the intercept probability analyzed in our previous work [22]. Technically speaking, it is much more challenging to obtain a closed-form expression of the secrecy outage probability than that of the intercept probability for cognitive radio networks, especially in the presence of coordinated eavesdroppers.

The following summarizes the main contributions of this paper. First, we propose the round-robin scheduling, optimal user scheduling and suboptimal user scheduling to protect the cognitive transmissions against the uncoordinated and coordinated eavesdroppers. The difference between the optimal and suboptimal scheduling schemes lies in that the optimal scheduling assumes the perfect CSI of all links from CUs to CBS, to PU and to eavesdroppers, whereas the suboptimal scheduling only needs the CSI of CUs-CBS links. Since the PU's and eavesdroppers' CSI is challenging to obtain at CUs in practical systems, the suboptimal user scheduling scheme is more attractive than the optimal user scheduling from this perspective, although the latter scheme may achieve a better security performance. Second, we derive closed-form expressions of the secrecy outage probability for the round-robin scheduling as well as the optimal and suboptimal user scheduling schemes with a PU's QoS constraint for both the uncoordinated and coordinated eavesdroppers. Last, we characterize the secrecy diversity orders of these three schemes through an asymptotic secrecy outage analysis and show that no matter whether the eavesdroppers collaborate or not, the round-robin scheduling achieves the diversity order of only one, whereas the optimal and suboptimal user scheduling schemes obtain the diversity order of $M$, where $M$ is the number of CUs.

The remainder of this paper is organized as follows. We first present the system model of a MUME cognitive radio network in Section II. Then, Section III proposes the round-robin scheduling, the optimal user scheduling and the suboptimal user scheduling in the presence of {multiple uncoordinated and coordinated eavesdroppers}. The closed-form secrecy outage expressions of various user scheduling schemes are also derived {for both the uncoordinated and coordinated eavesdroppers}. Next, in Section IV, we carry out the secrecy diversity analysis of the round-robin scheduling as well as the optimal and suboptimal scheduling schemes, followed by Section V, where numerical secrecy outage results of these three schemes are provided. Finally, some concluding remarks are drawn in Section VI.

\section{System Model and Problem Formulation}
\begin{figure}
  \centering
  {\includegraphics[scale=0.58]{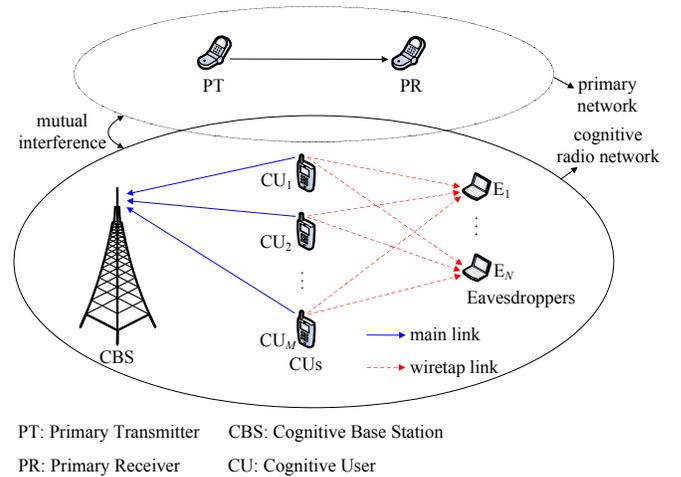}\\
  \caption{A multi-user multi-eavesdropper (MUME) cognitive radio network coexists with a primary network.}\label{Fig1}}
\end{figure}
As shown in Fig. 1, a multi-user multi-eavesdropper cognitive radio network consisting of one CBS, $M$ CUs and $N$ eavesdroppers shares the spectrum that is licensed to a primary network including one PT and one PR. Throughout this paper, we consider the use of underlay spectrum sharing, i.e., a CU and a PT are allowed to transmit simultaneously over the same spectrum, as long as the interference caused by CU is tolerable at PR and the quality of service (QoS) of PT-PR transmission is unaffected. {We consider that PT transmits to PR without power control and a maximum interference power $I$ is assumed to be tolerable at PR without affecting its QoS. This means that the interference received at PR from $\textrm{CU}_i$ must be less than the maximum tolerable level $I$ for the sake of protecting the primary QoS.} Considering that $\textrm{CU}_i$ transmits to CBS over the same spectrum band as the PT, we shall limit the transmit power of $\textrm{CU}_i$ denoted by $P_i$ as
\begin{equation}\label{equa1}
{P_i} = \frac{I}{{|{h_{ip}}{|^2}}},
\end{equation}
where ${h_{ip}}$ represents the fading coefficient of the $\textrm{CU}_i$-PR channel. {It is pointed out that the simple power control model given by (1) is widely used in literature [23]-[25] for characterizing the underlay cognitive radio. As shown in (1), the transmit power of $\textrm{CU}_i$ is a function of the random fading $|{h_{ip}}|^2$, which makes the closed-form secrecy outage probability analysis of the cognitive transmissions become more challenging. Considering a maximum power constraint $P$, the transmit power of $\textrm{CU}_i$ may be modeled as $P_i  = \min (\frac{I}{{|h_{ip} |^2 }},P)$. Since only a constant $P$ is introduced in this model, it will not result in any additional challenges in the secrecy outage analysis and, moreover, no new insight into the secrecy outage probability will be provided, as compared to the power control model of (1). Thus, we consider the use of (1) in modeling the $\textrm{CU}_i$'s transmit power throughout this paper.} In Fig. 1, $M$ CUs transmit their data packets to CBS, which is a typical uplink transmission scenario in cognitive radio networks [2]. Meanwhile, there are $N$ eavesdroppers in the cognitive radio network, which attempt to intercept the packets transmitted from CUs to CBS. For notational convenience, we denote $M$ CUs and $N$ eavesdroppers by ${\cal{U}}=\{{\textrm{CU}}_i|i=1,2,\cdots,M\}$ and ${\cal{E}}=\{{\textrm{E}}_j|j=1,2,\cdots,N\}$, respectively.

In addition, when a CU and a PT simultaneously transmit to their respective destination nodes, PT also causes an interference to CBS in decoding the CU's signal. {Following [23] and [26], the interference received at CBS from PT is considered to be a complex Gaussian random variable under an assumption that the primary signal may be generated by the random Gaussian codebook.} Moreover, the thermal noise at CBS is also complex Gaussian distributed. Thus, the interference plus noise at CBS, denoted by $n_b$, can be modeled as a complex Gaussian random variable with zero mean and variance $N_b$, which is represented by $n_b \sim {\mathcal{CN}}(0,N_b)$. Similarly, we can also model the interference plus noise received at an eavesdropper ${\textrm{E}}_j$, denoted by $n_{e_j}$, as a complex Gaussian random variable i.e. $n_{e_j} \sim {\mathcal{CN}}(0,N_{e_j})$. In the cognitive radio network shown in Fig. 1, $M$ CUs may access the licensed band and transmit to CBS using an orthogonal multiple access method e.g. the orthogonal frequency division multiple access (OFDMA). Generally speaking, the licensed band is first divided into multiple subchannels which are then assigned to $M$ CUs. Given a subchannel, we may need to determine which CU should be selected to access the subchannel, which will be discussed in the following Section III. Without loss of generality, considering that $\textrm{CU}_i$ transmits its signal $x_i$ to CBS with power $P_i$, we can express the received signal at CBS as
\begin{equation}\label{equa2}
{y_{ib}} = \sqrt {\frac{I}{{|{h_{ip}}{|^2}}}} {h_{ib}}{x_i} + {n_b},
\end{equation}
where ${h_{ib}}$ is the fading coefficient of the channel from
$\textrm{CU}_i$ to CBS and $n_b \sim {\mathcal{CN}}(0,N_b)$ represents the interference and thermal noise received at CBS. Using the Shannon's channel capacity formula, the capacity of
the main channel from $\textrm{CU}_i$ to CBS can be obtained from (2) as
\begin{equation}\label{equa3}
{C_{ib}} = {\log _2}(1 +
\frac{{I|{h_{ib}}{|^2}}}{{|{h_{ip}}{|^2}{N_b}}}).
\end{equation}
Meanwhile, due to the broadcast nature of radio propagation, the $\textrm{CU}_i$-CBS transmission may also be overheard by $N$ eavesdroppers. Thus, the signal received at an eavesdropper ${\textrm{E}}_j$ can be written as
\begin{equation}\label{equa4}
{y_{ie_j}} = \sqrt{\frac{I}{{|{h_{ip}}{|^2}}}} {h_{ie_j}}{x_i} + {n_{e_j}},
\end{equation}
where ${h_{ie_j}}$ is the fading coefficient of the channel from $\textrm{CU}_i$ to ${\textrm{E}}_j$ and $n_{e_j} \sim {\mathcal{CN}}(0,N_{e_j})$ represents the interference and thermal noise received at eavesdropper ${\textrm{E}}_j$. Similarly to (3), the capacity of the wiretap channel from $\textrm{CU}_i$ to ${\textrm{E}}_j$ is obtained from (4) as
\begin{equation}\label{equa5}
{C_{ie_j}} = {\log _2}(1 + \frac{{I|{h_{ie_j}}{|^2}}}{{|{h_{ip}}{|^2}{N_{e_j}}}}).
\end{equation}

{In this paper, we consider two eavesdropping scenarios: 1) uncoordinated case, where the eavesdroppers are independent of each other in intercepting the $\textrm{CU}_i$-CBS transmission; and 2) coordinated case, where the eavesdroppers collaborate for intercepting the cognitive transmissions.} In the uncoordinated case, the eavesdroppers perform the interception independently and the $\textrm{CU}_i$-CBS transmission is secure when all $N$ eavesdroppers fail to decode the signal $x_i$. Thus, the overall capacity of the wiretap channel from  $\textrm{CU}_i$ to $N$ eavesdroppers can be given by the maximum of individual achievable rates at $N$ eavesdroppers, yielding
\begin{equation}\label{equa6}
{C_{ie}}=\mathop {\max }\limits_{e_j \in {\cal{E}}} {C_{{ie_j}}} =  \mathop {\max }\limits_{e_j \in {\cal{E}}} {\log _2}(1 +
\frac{{I|{h_{ie_j}}{|^2}}}{{|{h_{ip}}{|^2}{N_{e_j}}}}),
\end{equation}
for the uncoordinated case, where ${\cal E}$ denotes the set of $N$ eavesdroppers. {In the coordinated case, $N$ eavesdroppers first combine their received signals to obtain an enhanced version for the sake of improving the possibility of successfully decoding the signal $x_i$. Considering the maximal ratio combining (MRC) and using (4), we obtain a combined version of the received signals at $N$ eavesdroppers as
\begin{equation}\nonumber
y_{ie}  = \sum\limits_{e_j  \in {\cal{E}} } {\sqrt {\frac{I}{{|h_{ip} |^2 }}} |h_{ie_j } |^2 x_i }  + \sum\limits_{e_j  \in {\cal{E}} } {h_{ie_j }^* n_{e_j } },
\end{equation}
from which the overall capacity of the wiretap channel from  $\textrm{CU}_i$ to $N$ eavesdroppers is given by
\begin{equation}\label{equa7}
C_{ie}  = \log _2 [1 + \frac{{I(\sum\limits_{e_j  \in {\cal{E}} } {|h_{ie_j } |^2 } )^2 }}{{|h_{ip} |^2 \sum\limits_{e_j  \in {\cal{E}} } {(|h_{ie_j } |^2 N_{e_j } )} }}],
\end{equation}
for the coordinated case.} As discussed in [8] and [9], the secrecy capacity of wireless transmissions is shown as the difference between the capacity of the main channel and that of the wiretap channel. Thus, we can obtain the secrecy capacity of $\textrm{CU}_i$-CBS transmission in the presence of $N$ eavesdroppers as
\begin{equation}\label{equa8}
{C^s_i} = {C_{ib}} - C_{ie},
\end{equation}
where ${C_{ib}}$ is given by (3) and $C_{i{e}}$ is characterized by (6) and (7) {for the uncoordinated and coordinated cases, respectively. Additionally, all the wireless channels shown in Fig. 1 (i.e., $h_{ip}$, $h_{ib}$ and $h_{ie_j}$) are characterized with the Rayleigh fading model. The average channel gains of $|h_{ip}|^2$, $|h_{ib}|^2$ and $|h_{ie_j}|^2$ are denoted by $\sigma_{ip}^2$, $\sigma_{ib}^2$ and $\sigma_{ie_j}^2$, respectively. Moreover, although only the Rayleigh fading model is considered in this paper, similar performance analysis can be obtained for other fading channel models (e.g., Nakagami model).}

\section{Multi-user Scheduling Schemes and Secrecy Outage Analysis}
In this section, we present several multi-user scheduling schemes including the round-robin scheduling, the optimal user scheduling, and the suboptimal user scheduling {in the presence of the uncoordinated and coordinated eavesdroppers}. The optimal user scheduling is aimed to maximize the secrecy capacity of the cognitive transmissions from CUs to CBS, assuming that the CSIs of all CUs-CBS, CUs-PR, and CUs-${\textrm{E}}_j$ links are available. By contrast, the suboptimal user scheduling only assumes that the CSIs of CUs-CBS links are known, attempting to address the multi-user scheduling without the PR's and eavesdroppers' CSI knowledge. The closed-form secrecy outage probability expressions of the round-robin scheduling as well as the optimal and suboptimal scheduling are also derived {for both the uncoordinated and coordinated eavesdroppers}.

\subsection{Round-Robin Scheduling}
This subsection presents the conventional round-robin scheduling as a benchmark. With the round-robin scheduling, $M$ CUs take turns in accessing the licensed spectrum and thus each user has an equal chance to transmit its signal to CBS. As is known, a secrecy outage event occurs when the secrecy capacity drops below a predefined secrecy rate $R_s$. Thus, given that ${\textrm{CU}}_i$ transmits to CBS, the secrecy outage probability of ${\textrm{CU}}_i$-CBS transmission is obtained as
\begin{equation}\label{equa9}
{P_{out,i}} = \Pr \left( {C_i^s < {R_s}} \right),
\end{equation}
where $C_i^s$ is given by (8). Substituting (3) and (6) into (8) and combining with (9) yield
\begin{equation}\label{equa10}
\begin{split}
{P_{out,i}}
= \Pr \left( {\mathop {\max }\limits_{{e_j} \in {\cal {E}}} \frac{{|{h_{i{e_j}}}{|^2}}}{{{N_{{e_j}}}}} > \frac{1}{{{2^{{R_s}}}{N_b}}}|{h_{ib}}{|^2} - \frac{{{2^{{R_s}}} - 1}}{{{2^{{R_s}}}I}}|{h_{ip}}{|^2}} \right),
\end{split}
\end{equation}
{for the uncoordinated case}, which is further obtained as (see Appendix A)
\begin{equation}\label{equa11}
{P_{out,i}} = \frac{{\sigma _{ip}^2({2^{{R_s}}} - 1){N_b} + \sum\limits_{n = 1}^{{2^N} - 1} {\frac{{{{( - 1)}^{|{{\cal {E}}_n}| + 1}}{2^{{R_s}}}{N_b}I}}{{\sigma _{ib}^{ - 2}{2^{{R_s}}}{N_b} + \sum\limits_{{e_j} \in {{\cal {E}}_n}} {{{(\sigma _{i{e_j}}^{-2}{N_{{e_j}}})}}} }}} }}{{\sigma _{ib}^2I + \sigma _{ip}^2({2^{{R_s}}} - 1){N_b}}},
\end{equation}
where $N$ is the number of eavesdroppers, ${\cal {E}}_n$ represents the $n$-th non-empty subset of the elements of ${\cal {E}}$, and $|{\cal {E}}_n|$ is the cardinality of set ${\cal {E}}_n$. {Additionally, it is observed from (7) that obtaining a general closed-form expression of the secrecy outage probability ${P_{out,i}}$ for the coordinated case is challenging. For simplicity, we assume that the fading coefficients of all ${\textrm{CU}}_i$-${\textrm{E}}_j$ channels $|h_{ie_j}|^2$ are independent and identically distributed (i.i.d.) random variables for different eavesdroppers with the same average channel gain denoted by $\sigma^2_{ie}=E(|h_{ie_j}|^2)$. This assumption is widely used and valid in a statistical sense when all eavesdroppers are uniformly distributed around CUs. Moreover, proceeding as Appendix A and assuming that different eavesdroppers have the same noise variance of $N_{e_j}=N_e$, we can obtain the secrecy outage probability ${P_{out,i}}$ from (7), (8) and (9) as
\begin{equation}\label{equa12}
P_{out,i}  = 1 - \frac{{\sigma _{ib}^2 I}}{{\sigma _{ib}^2 I + \sigma _{ip}^2 (2^{R_s}  - 1)N_b }}(1 + \frac{{2^{R_s} \sigma _{ie}^2 N_b }}{{\sigma _{ib}^2 N_e }})^{ - N},
\end{equation}
for the coordinated case.} As aforementioned, the round-robin scheduling scheme allows $M$ CUs to take turns in accessing the licensed spectrum and thus the secrecy outage probability of the round-robin scheduling is the mean of $M$ CUs' secrecy outage probabilities, yielding
\begin{equation}\label{equa13}
P_{out}^{round} = \frac{1}{M}\sum\limits_{i = 1}^M {{P_{out,i}}},
\end{equation}
where $M$ is the number of CUs and ${P_{out,i}}$ is given by (11) and (12) {for the uncoordinated and coordinated cases, respectively.}

\subsection{Optimal User Scheduling}
In this subsection, we propose an optimal user scheduling scheme for the sake of improving the security of the CUs-CBS transmissions. Considering $M$ CUs available in the cognitive radio network, a CU with the highest secrecy capacity is selected to access a given spectrum band. Therefore, using (8), we can express the optimal user scheduling criterion as
\begin{equation}\label{equa14}
{\textrm{Optimal User}} = \arg \mathop {\max }\limits_{i \in {\cal {U}}}  {C^s_i} ,
\end{equation}
where ${\cal {U}}$ represents the set of CUs. Substituting (3) and (6) into (14) gives
\begin{equation}\label{equa15}
{\textrm{Optimal User}} = \arg \mathop {\max }\limits_{i \in {\cal {U}}} \left( {\frac{{|{h_{ip}}{|^2} + I {{|{h_{ib}}{|^2}}}{N^{-1}_b}}}{{|{h_{ip}}{|^2} + I\mathop {\max }\limits_{{e_j} \in {\cal {E}}} {{|{h_{i{e_j}}}{|^2}}}N^{-1}_{{e_j}}}}} \right),
\end{equation}
{for the uncoordinated case.} Moreover, substituting (3) and (7) into (14) yields
\begin{equation}\label{equa16}
\begin{split}
{\textrm{Optimal User}} = \arg \mathop {\max }\limits_{i \in {\cal {U}}} \left( \frac{{|h_{ip} |^2  + I|h_{ib} |^2 N_b^{ - 1} }}{{|h_{ip} |^2  + \frac{I(\sum\limits_{e_j  \in {\cal E}} {|h_{ie_j } |^2 } )^2} {\sum\limits_{e_j  \in {\cal E}} {(|h_{ie_j } |^2 N_{e_j } )}} }} \right),
\end{split}
\end{equation}
{for the coordinated case.} {One can observe from (15) and (16) that the CSIs ${|{h_{ib}}{|^2}}$, ${|{h_{ip}}{|^2}}$ and $|{h_{i{e_j}}}|^2$ of the ${\textrm{CU}}_i$-CBS, ${\textrm{CU}}_i$-PR and ${\textrm{CU}}_i$-${\textrm{E}}_j$ links as well as the number of eavesdroppers $N$ and the noise variance $N_{e_j}$ are assumed in determining the optimal user among $M$ CUs. However, the PR's and eavesdroppers' CSIs, the number of eavesdroppers, and the noise variance may be unavailable in some cases. To this end, the following subsection will consider the multi-user scheduling without the need of these information.} Using (14), the secrecy outage probability of the proposed optimal user scheduling scheme can be obtained as
\begin{equation}\label{equa17}
\begin{split}
P_{out}^{optimal} = \Pr \left( {\mathop {\max }\limits_{i \in {\cal {U}}} {C^s_i} < {R_s}} \right)= \prod\limits_{i \in {\cal {U}}} {\Pr \left( {C_i^s < {R_s}} \right)},  \\
 \end{split}
\end{equation}
where $C^s_i$ is given by (8). Combining (9) and (17), we obtain the secrecy outage probability of the optimal scheduling as
\begin{equation}\label{equa18}
P_{out}^{optimal} = \prod\limits_{i \in {\cal{U}}} {P_{out,i}},
\end{equation}
where $P_{out,i}$ is given by (11) and (12) {for the uncoordinated and coordinated cases, respectively.}

\subsection{Suboptimal User Scheduling}
This subsection proposes a suboptimal user scheduling scheme under the condition that only the CSIs of CUs-CBS channels are available without knowing the CSI knowledge of the primary receiver and eavesdroppers. Since only the CSIs of CUs-CBS channels are known in this case, a CU with the highest instantaneous fading gain to CBS is typically regarded as the optimal user, yielding
\begin{equation}\label{equa19}
\begin{split}
{\textrm{Optimal User}} =  \arg \mathop {\max }\limits_{i \in {\cal {U}}} {|{h_{ib}}{|^2}},
\end{split}
\end{equation}
where ${\cal {U}}$ represents the set of $M$ CUs. It is observed from (19) that only $|h_{ib}|^2$ is needed in the suboptimal user scheduling scheme without the PR's and eavesdroppers' CSIs $|h_{ip}|^2$ and $|h_{ie_j}|^2$. This is different from the aforementioned optimal user scheduling scheme which requires the CSIs of all ${\textrm{CU}}_i$-CBS, ${\textrm{CU}}_i$-PR and ${\textrm{CU}}_i$-${\textrm{E}}_j$ channels (i.e., $|h_{ib}|^2$, $|h_{ip}|^2$, and $|h_{ie_j}|^2$). For notational convenience, let `$o$' denote the optimal user determined by (19). Thus, the secrecy capacity of the transmission from the optimal user ($o$) to CBS in the presence of $N$ {uncoordinated eavesdroppers} is obtained as
\begin{equation}\label{equa20}
C_o^s = {C_{ob}} - \mathop {\max }\limits_{{e_j} \in {\cal {E}}} {C_{o{e_j}}},
\end{equation}
where ${C_{ob}}$ and ${C_{o{e_j}}}$, respectively, represent the channel capacities from the optimal user to CBS and to eavesdropper ${\textrm{E}}_j$, which are given by
\begin{equation}\label{equa21}
{C_{ob}} = {\log _2}(1 + \frac{{|{h_{ob}}{|^2}I}}{{|{h_{op}}{|^2}{N_b}}}),
\end{equation}
and
\begin{equation}\label{equa22}
{C_{o{e_j}}} = {\log _2}(1 + \frac{{|{h_{o{e_j}}}{|^2}I}}{{|{h_{op}}{|^2}{N_{e_j}}}}),
\end{equation}
where $|h_{ob}|^2$, $|h_{op}|^2$, and $|h_{o{e_j}}|^2$ represent fading coefficients of the channels from the optimal user to CBS, to PR, and to eavesdropper ${\textrm{E}}_j$, respectively. Combining (20)-(22), we obtain the secrecy outage probability of the proposed suboptimal user scheduling scheme as
\begin{equation}\label{equa23}
\begin{split}
&P_{out}^{sub} = \Pr \left( {C_o^s < {R_s}} \right) \\
&= \Pr \left( {{2^{{R_s}}}I\mathop {\max }\limits_{{e_j} \in  {\cal {E}}} \frac{{|{h_{o{e_j}}}{|^2}}}{{{N_{{e_j}}}}} > \frac{I}{{{N_b}}}|{h_{ob}}{|^2} - ({2^{{R_s}}} - 1)|{h_{op}}{|^2}} \right), \\
 \end{split}
\end{equation}
{for the uncoordinated case.} By using the law of total probability and denoting $t=\frac{I}{{{N_b}}}|{h_{ib}}{|^2} - ({2^{{R_s}}} - 1)|{h_{ip}}{|^2}$, (23) is rewritten as
\begin{equation}\label{equa24}
P_{out}^{sub} = \sum\limits_{i = 1}^M {\Pr \left( {{2^{{R_s}}}I\mathop {\max }\limits_{{e_j} \in {\cal {E}}} \frac{{|{h_{i{e_j}}}{|^2}}}{{{N_{{e_j}}}}} > t,{\textrm{ }} o = i} \right)}.
\end{equation}
Combining (19) and (24), we have
\begin{equation}\label{equa25}
P_{out}^{sub} = \sum\limits_{i = 1}^M {\Pr \left(
\begin{split}
&{{2^{{R_s}}}I\mathop {\max }\limits_{{e_j} \in {\cal{E}}} \frac{{|{h_{i{e_j}}}{|^2}}}{{{N_{{e_j}}}}} >t,}\\
&{\textrm{ }}\mathop {\max }\limits_{\scriptstyle k \in {\cal{U}} \hfill \atop\scriptstyle k \ne i \hfill} |{h_{kb}}{|^2} < |{h_{ib}}{|^2}
\end{split}
\right)}.
\end{equation}
By using the result of Appendix B, the secrecy outage probability $P_{out}^{sub}$ is obtained from (25) as
\begin{equation}\label{equa26}
\begin{split}
P_{out}^{sub} =& \sum\limits_{i = 1}^M {\sum\limits_{n = 1}^{{2^N} - 1} {{{( - 1)}^{|{{\cal{E}}_n}| + 1}}\left( {P_{out,I}^{sub} - P_{out,II}^{sub}} \right)} } \\
&+ \sum\limits_{i = 1}^M {P_{out,III}^{sub}} ,
\end{split}
\end{equation}
{for the uncoordinated case}, where ${\cal {E}}_n$ represents the $n$-th non-empty subset of the elements of ${\cal {E}}$, $P_{out,I}^{sub}$, $P_{out,II}^{sub}$ and $P_{out,III}^{sub}$ are given by (B.11)-(B.12), (B.13)-(B.14) and (B.15) respectively. {The following presents the secrecy outage probability analysis of the suboptimal user scheduling for the coordinated eavesdroppers. As mentioned earlier in Section III-A, it is challenging to obtain a general closed-form expression of the secrecy outage probability for the coordinated case. We here consider that the fading coefficients $|h_{ie_j}|^2$ for $e_j \in {\cal E}$ are i.i.d. with the same mean of $\sigma^2_{ie}$ and different eavesdroppers have the same noise variance of $N_e$. Hence, substituting $N_{e_j}=N_e$ into (7) and using the law of total probability, we obtain the secrecy outage probability $P_{out}^{sub}$ of the suboptimal user scheduling as (27)
\begin{figure*}
\begin{equation}\label{equa27}
P_{out}^{sub} = \sum\limits_{i = 1}^M {\Pr \left( {\sum\limits_{e_j  \in {\cal E}} {|h_{ie_j } |^2 }  > \frac{{N_e }}{{2^{R_s } N_b }}|h_{ib} |^2  - \frac{{(2^{R_s }  - 1)N_e }}{{2^{R_s } I}}|h_{ip} |^2 ,\mathop {\max }\limits_{\scriptstyle k \in {\cal U} \hfill \atop
  \scriptstyle k \ne i \hfill} |h_{kb} |^2  < |h_{ib} |^2 } \right)}
\end{equation}
\end{figure*}
for the coordinated case, which is further given by (see Appendix C)
\begin{equation}\label{equa28}
P_{out}^{sub}  = \sum\limits_{i = 1}^M {(P_{out,I}  + P_{out,II}  + P_{out,III} )},
\end{equation}
where $P_{out,I}$, $P_{out,II}$ and $P_{out,III}$ are given by (C.13), (C.14) and (C.15), respectively. So far, we have derived closed-form secrecy outage expressions for the round-robin scheduling as well as the optimal and suboptimal scheduling schemes in the presence of the uncoordinated and coordinated eavesdroppers, which will be used in Section V for conducting numerical evaluation of the secrecy outage performance.}

\section{Secrecy Diversity Analysis}
In this section, we analyze the secrecy diversity performance of multi-user cognitive transmissions in the presence of multiple {uncoordinated and coordinated eavesdroppers}. Although the closed-form secrecy outage expressions shown in (13), (18), (26) and (28) can be used to show the transmission security performance of various user scheduling schemes, they fail to provide an intuitive insight into the impact of the number of CUs and eavesdroppers on the cognitive transmission security. As a consequence, this section presents the secrecy diversity analysis of the round-robin scheduling as well as the optimal and suboptimal scheduling schemes.

\subsection{Round-Robin Scheduling}
Let us consider the round-robin scheduling as a baseline for comparison. First, the cognitive radio transmission is subject to the primary QoS constraint i.e. the maximum tolerable interference level at PR $I$. Generally speaking, with an increasing $I$, the secrecy outage probability of cognitive transmissions decreases accordingly. From (11) and (13), we obtain
\begin{equation}\label{equa29}
\mathop {\lim }\limits_{I \to \infty } P_{out}^{round} = \frac{1}{M}\sum\limits_{i = 1}^M {\sum\limits_{n = 1}^{{2^N} - 1} {\frac{{{{( - 1)}^{|{{\cal {E}}_n}| + 1}}{2^{{R_s}}}{N_b}}}{{{2^{{R_s}}}{N_b} + \sum\limits_{{e_j} \in {{\cal {E}}_n}} {\sigma _{ib}^2\sigma _{i{e_j}}^{ - 2}N_{{e_j}}} }}} } ,
\end{equation}
{for the uncoordinated case.}
One can observe from (29) that as the maximum tolerable interference level $I$ tends to infinity, the secrecy outage probability of the round-robin scheduling scheme converges to a non-zero constant. From (1), an infinite $I$ means that the transmit power of CUs approaches infinity. Hence, as the CUs' transmit power increases to infinity, a secrecy outage probability floor occurs. {Notice that the secrecy outage floor provides a lower bound on the secrecy outage probability that a cognitive radio system can achieve with high interference temperature. It is also meaningful and effective to employ the secrecy outage floor as a metric to evaluate the security performance of different signal processing techniques in a cognitive radio system.} For notational convenience, the secrecy outage floor of the round-robin scheduling scheme is denoted by $P_{out,floor}^{round}$, i.e., $P_{out,floor}^{round} = \mathop {\lim }\limits_{I \to \infty } P_{out}^{round}$. Denoting $\sigma_{ib}^2=\theta_{ib}\sigma^2_{m}$ and $\sigma_{ie_j}^2=\theta_{ie_j}\sigma^2_{e}$, where $\sigma^2_{m}$ and $\sigma^2_{e}$, respectively, represent the reference channel gain of the main links from CUs to CBS and that of the wiretap links from CUs to eavesdroppers, we may obtain the secrecy outage floor of the round-robin scheduling scheme from (29) as
\begin{equation}\label{equa30}
P_{out,floor}^{round} = \frac{1}{M}\sum\limits_{i = 1}^M {\sum\limits_{n = 1}^{{2^N} - 1} {\frac{{{{( - 1)}^{|{{\cal {E}}_n}| + 1}}{2^{{R_s}}}{N_b}}}{{{2^{{R_s}}}{N_b} + {\lambda _{me}
}\sum\limits_{{e_j} \in {{\cal {E}}_n}} {{\theta _{ib}}\theta _{i{e_j}}^{ - 1}N_{{e_j}}} }}} },
\end{equation}
{for the uncoordinated case}, where $\lambda_{me}=\sigma^2_{m}/\sigma^2_{e}$ is called the main-to-eavesdropper ratio (MER). {The traditional diversity gain is defined in [27] as
\begin{equation}\nonumber
d =  - \mathop {\lim }\limits_{{\textrm{SNR}} \to \infty } \frac{{\log P_e ({\textrm{SNR}})}}{{\log {\textrm{SNR}}}},
\end{equation}
where SNR stands for the signal-to-noise ratio and ${P_e ({\textrm{SNR}})}$ represents the bit error rate as a function of SNR. However, as the CUs' transmit power increases to infinity, the secrecy outage probability of (30) tends to a non-zero constant, which makes the traditional diversity definition become inappropriate for the secrecy outage analysis. It is also observed from (30) that with an infinite transmit power, the secrecy outage probability becomes nothing to do with the CUs-PR channel $h_{ip}$ and is mainly determined by the main channel $h_{ib}$ and eavesdropping channel $h_{ie_j}$. Motivated by this observation, we here define a secrecy diversity gain as an asymptotic ratio of the logarithmic secrecy outage floor to the logarithmic MER $\lambda_{me}$ (i.e., the ratio between the reference gains of the main channel and eavesdropping channel) as $\lambda_{me} \to \infty$ [28], yielding
\begin{equation}\label{equa31}
{d_{round}} =  - \mathop {\lim }\limits_{{\lambda _{me}} \to \infty } \frac{{\log (P_{out,floor}^{round})}}{{\log ({\lambda _{me}})}},
\end{equation}
which, in turn, results in the secrecy outage floor $P_{out,floor}^{round}$ behaving as $ \lambda _{me}^{ - d_{round} }$ in high MER region. This also shows that as MER increases, the secrecy outage floor $P_{out,floor}^{round}$ decreases faster with a higher diversity order $d_{round}$. Therefore, the secrecy diversity order can be used as a simple but effective metric to evaluate the secrecy outage floor performance, especially in high MER region.} Substituting (30) into (31), we obtain the secrecy diversity of the round-robin scheduling scheme as
\begin{equation}\label{equa32}
{d_{round}} =  1,
\end{equation}
{for the uncoordinated case, which shows that the secrecy diversity order of only one is achieved by the round-robin scheduling scheme, when the eavesdroppers are independent of each other in intercepting the cognitive transmissions. In what follows, we analyze the secrecy diversity of the round-robin scheduling for the coordinated case. Noting $N_{e_j}>0$ and using the inequality $\mathop {\min }\limits_{e_j  \in {\cal E}} N_{e_j } \sum\limits_{e_j  \in {\cal E}} {|h_{ie_j } |^2 }  \le \sum\limits_{e_j  \in {\cal E}} {|h_{ie_j } |^2 N_{e_j } }  \le \mathop {\max }\limits_{e_j  \in {\cal E}} N_{e_j } \sum\limits_{e_j  \in {\cal E}} {|h_{ie_j } |^2 }$ into (7), we have
\begin{equation}\label{equa33}
\log _2 (1 + \frac{{I\sum\limits_{e_j  \in {\cal E}} {|h_{ie_j } |^2 } }}{{|h_{ip} |^2 \mathop {\max }\limits_{e_j  \in {\cal E}} N_{e_j } }}) \le C_{ie}  \le \log _2 (1 + \frac{{I\sum\limits_{e_j  \in {\cal E}} {|h_{ie_j } |^2 } }}{{|h_{ip} |^2 \mathop {\min }\limits_{e_j  \in {\cal E}} N_{e_j } }}),
\end{equation}
which may be further given by
\begin{equation}\label{equa34}
\log _2 (1 + \frac{{I\mathop {\max }\limits_{e_j  \in {\cal E}} |h_{ie_j } |^2 }}{{|h_{ip} |^2 \mathop {\max }\limits_{e_j  \in {\cal E}} N_{e_j } }}) \le C_{ie}  \le \log _2 (1 + \frac{{NI\mathop {\max }\limits_{e_j  \in {\cal E}} |h_{ie_j } |^2 }}{{|h_{ip} |^2 \mathop {\min }\limits_{e_j  \in {\cal E}} N_{e_j } }}),
\end{equation}
which is obtained by using the inequality $\mathop {\max }\limits_{e_j  \in {\cal E}} |h_{ie_j } |^2  \le \sum\limits_{e_j  \in {\cal E}} {|h_{ie_j } |^2 }  \le N\mathop {\max }\limits_{e_j  \in {\cal E}} |h_{ie_j } |^2$, where $N$ is the number of eavesdroppers. Combining (33) and (34) with (9), we have
\begin{equation}\label{equa35}
P_{out,i}^{lower}  \le P_{out,i}  \le P_{out,i}^{upper},
\end{equation}
where the lower and upper bounds $P_{out,i}^{lower}$ and $P_{out,i}^{upper}$ are given by
\begin{equation}\label{equa36}
P_{out,i}^{lower}  = \Pr \left( {
\begin{split}
&\mathop {\max }\limits_{e_j  \in {\cal E}} \frac{{|h_{ie_j } |^2 }}{{\mathop {\max }\limits_{e_j  \in {\cal E}} N_{e_j } }} > \frac{1}{{2^{R_s } N_b }}|h_{ib} |^2  \\
&\quad\quad\quad\quad\quad\quad- \frac{{2^{R_s }  - 1}}{{2^{R_s } I}}|h_{ip} |^2
\end{split}
} \right),
\end{equation}
and
\begin{equation}\label{equa37}
P_{out,i}^{upper}  = \Pr \left( {
\begin{split}
&N\mathop {\max }\limits_{e_j  \in {\cal E}} \frac{{|h_{ie_j } |^2 }}{{\mathop {\min }\limits_{e_j  \in {\cal E}} N_{e_j } }} > \frac{1}{{2^{R_s } N_b }}|h_{ib} |^2  \\
&\quad\quad\quad\quad\quad\quad\quad - \frac{{2^{R_s }  - 1}}{{2^{R_s } I}}|h_{ip} |^2
\end{split}
} \right),
\end{equation}
for the coordinated case. Comparing (36) and (37) with (10) and using (30), we readily obtain
\begin{equation}\label{equa38}
\mathop {\lim }\limits_{I \to \infty } P_{out,i}^{lower}  = \sum\limits_{n = 1}^{2^N  - 1} {\frac{{( - 1)^{|{\cal E}_n | + 1} 2^{R_s } N_b }}{{2^{R_s } N_b  + \lambda _{me} \sum\limits_{e_j  \in {\cal E}_n } {\theta _{ib} \theta _{ie_j }^{ - 1} \mathop {\max }\limits_{e_j  \in {\cal E}} N_{e_j }  } }}},
\end{equation}
and
\begin{equation}\label{equa39}
\mathop {\lim }\limits_{I \to \infty } P_{out,i}^{upper}  = \sum\limits_{n = 1}^{2^N  - 1} {\frac{{( - 1)^{|{\cal E}_n | + 1} 2^{R_s } N_b }}{{2^{R_s } N_b  + \lambda _{me} \sum\limits_{e_j  \in {\cal E}_n } {\theta _{ib} \theta _{ie_j }^{ - 1} N^{-1}\mathop {\min }\limits_{e_j  \in {\cal E}} N_{e_j }  } }}}.
\end{equation}
Combining (38) and (39) with (13) yields
\begin{equation}\label{equa40}
\frac{1}{M}\sum\limits_{i = 1}^M {\mathop {\lim }\limits_{I \to \infty } P_{out,i}^{lower} }  \le P_{out,floor}^{round}  \le \frac{1}{M}\sum\limits_{i = 1}^M {\mathop {\lim }\limits_{I \to \infty } P_{out,i}^{upper} },
\end{equation}
for the coordinated case. Substituting (40) into (31) and using (38) and (39) give
\begin{equation}\label{equa41}
1 \le d_{round}  \le 1,
\end{equation}
which can be further obtained from the squeeze theorem as
\begin{equation}\label{equa42}
d_{round} =1,
\end{equation}
for the coordinated case. As shown in (32) and (42), no matter whether the eavesdroppers collaborate or not, the round-robin scheduling scheme always achieves the diversity order of only one. This also means that the round-robin scheme fails to achieve any secrecy diversity benefits with multiple CUs.}

\subsection{Optimal User Scheduling}
This subsection analyzes the secrecy diversity order of the proposed optimal user scheduling scheme. Using (18) and letting ${I \to \infty }$, we obtain the secrecy outage floor of the optimal user scheduling scheme as
\begin{equation}\label{equa43}
P_{out,floor}^{optimal} = \prod\limits_{i \in {\cal {U}}} {\mathop {\lim }\limits_{I \to \infty } {P_{out,i}}},
\end{equation}
where $\mathop {\lim }\limits_{I \to \infty } {P_{out,i}}$ is further computed from (11) as
\begin{equation}\label{equa44}
\mathop {\lim }\limits_{I \to \infty } {P_{out,i}} = \sum\limits_{n = 1}^{{2^N} - 1} {\frac{{{{( - 1)}^{|{{\cal {E}}_n}| + 1}}{2^{{R_s}}}{N_b}}}{{{2^{{R_s}}}{N_b} + {\lambda _{me}}\sum\limits_{{e_j} \in {{\cal {E}}_n}} {{\theta _{ib}}\theta _{i{e_j}}^{ - 1}N_{{e_j}}} }}},
\end{equation}
{for the uncoordinated case}, where $N$ is the number of eavesdroppers, $\theta_{ib}=\sigma_{ib}^2/\sigma^2_{m}$, $\theta_{ie_j}=\sigma_{ie_j}^2/\sigma^2_{e}$, and $\lambda_{me}=\sigma^2_{m}/\sigma^2_{e}$. Substituting (44) into (43) gives
\begin{equation}\label{equa45}
\begin{split}
 P_{out,floor}^{optimal}
=& \prod\limits_{i \in {\cal {U}}} {\left[ {\sum\limits_{n = 1}^{{2^N} - 1} {\frac{{{{( - 1)}^{|{{\cal {E}}_n}| + 1}}{2^{{R_s}}}{N_b}}}{{{2^{{R_s}}}{N_b}\lambda _{me}^{ - 1} + \sum\limits_{{e_j} \in {{\cal {E}}_n}} {{\theta _{ib}}\theta _{i{e_j}}^{ - 1}N_{{e_j}}} }}} } \right]}\\
& \quad \cdot {\left( {\frac{1}{{{\lambda _{me}}}}} \right)^M},
\end{split}
\end{equation}
where $M$ is the number of CUs. Similarly to (31), the secrecy diversity order of the optimal user scheduling scheme is defined as
\begin{equation}\label{equa46}
{d_{optimal}} =  - \mathop {\lim }\limits_{{\lambda _{me}} \to \infty } \frac{{\log (P_{out,floor}^{optimal})}}{{\log ({\lambda _{me}})}}.
\end{equation}
Substituting (45) into (46) yields
\begin{equation}\label{equa47}
{d_{optimal}} = M,
\end{equation}
{for the uncoordinated case, which demonstrates that the secrecy diversity order of $M$ is achieved by the optimal scheduling scheme when the eavesdroppers are independent of each other in tapping the cognitive transmissions. Similarly to (40), we may obtain the secrecy outage floor of the optimal user scheduling as
\begin{equation}\label{equa48}
\prod\limits_{i \in {\cal U}} {\mathop {\lim }\limits_{I \to \infty } P_{out,i}^{lower} }  \le P_{out,floor}^{optimal}  \le \prod\limits_{i \in {\cal U}} {\mathop {\lim }\limits_{I \to \infty } P_{out,i}^{upper} },
\end{equation}
for the coordinated case, where ${\mathop {\lim }\limits_{I \to \infty } P_{out,i}^{lower} }$ and ${\mathop {\lim }\limits_{I \to \infty } P_{out,i}^{upper} }$ are given by (38) and (39), respectively. Substituting (48) into (46), we have
\begin{equation}\label{equa49}
M \le d_{optimal}  \le M,
\end{equation}
from which the secrecy diversity of the optimal user scheduling scheme is readily obtained as
\begin{equation}\label{equa50}
d_{optimal}  = M,
\end{equation}
for the coordinated case. It is seen from (47) and (50) that for both the uncoordinated and coordinated eavesdroppers, the optimal user scheduling achieves the diversity order of $M$. This can also be interpreted as that the secrecy outage probability floor of the optimal user scheduling behaves as ${(\frac{1}{{{\lambda _{me}}}})^M}$ in high MER region. Therefore, with an increasing number of CUs, the secrecy outage floor of the optimal user scheduling decreases significantly, showing its advantage over the round-robin scheduling scheme.}

\subsection{Suboptimal User Scheduling}
This subsection is focused on the secrecy diversity analysis of the suboptimal user scheduling scheme. Let us first analyze the secrecy outage floor of the suboptimal user scheduling with an infinite $I$. From (25), we obtain (51) at the top of the following page
\begin{figure*}
\begin{equation}\label{equa51}
P_{out,floor}^{sub} = \mathop {\lim }\limits_{I \to \infty } P_{out}^{sub} = \sum\limits_{i = 1}^M {\Pr \left( {{2^{{R_s}}}\mathop {\max }\limits_{{e_j} \in {\cal {E}}} \frac{{|{h_{i{e_j}}}{|^2}}}{{{N_{{e_j}}}}} > \frac{1}{{{N_b}}}|{h_{ib}}{|^2},{\textrm{ }}\mathop {\max }\limits_{\scriptstyle k \in {\cal {U}} \hfill \atop
  \scriptstyle k \ne i \hfill} |{h_{kb}}{|^2} < |{h_{ib}}{|^2}} \right)}
\end{equation}
\end{figure*}
for the uncoordinated case. Considering that ${|{h_{i{e_j}}}{|^2}}$ and $|h_{kb}|^2$ are independent exponentially distributed random variables with respective means $\sigma^2_{ie_j}$ and $\sigma^2_{kb}$ and denoting $|{h_{ib}}{|^2} = x$, we can equivalently rewrite (51) as
\begin{equation}\label{equa52}
\begin{split}
 P_{out,floor}^{sub} = &\sum\limits_{i = 1}^M {\int_0^\infty  {\left[ {1 - \prod\limits_{{e_j} \in {\cal {E}}} {\left( {1 - \exp ( - \frac{{{N_{{e_j}}}x}}{{\sigma _{i{e_j}}^2{2^{{R_s}}}{N_b}}})} \right)} } \right]} }  \\
& \quad \times \prod\limits_{\scriptstyle k \in {\cal {U}} \hfill \atop
  \scriptstyle k \ne i \hfill} {\left( {1 - \exp ( - \frac{x}{{\sigma _{kb}^2}})} \right)} \frac{1}{{\sigma _{ib}^2}}\exp ( - \frac{x}{{\sigma _{ib}^2}})dx .
\end{split}
\end{equation}
Using the binomial theorem, $\prod\limits_{{e_j} \in {\cal {E}}} {\left( {1 - \exp ( - \frac{{{N_{{e_j}}}x}}{{\sigma _{i{e_j}}^2{2^{{R_s}}}{N_b}}})} \right)}$ can be expanded as
\begin{equation}\label{equa53}
\begin{split}
&\prod\limits_{{e_j} \in {\cal {E}}} {\left( {1 - \exp ( - \frac{{{N_{{e_j}}}x}}{{\sigma _{i{e_j}}^2{2^{{R_s}}}{N_b}}})} \right)} \\
&= 1 - \sum\limits_{n = 1}^{{2^N} - 1} {{{( - 1)}^{|{{\cal {E}}_n}| + 1}}\exp ( - \sum\limits_{{e_j} \in {{\cal {E}}_n}} {\frac{{{N_{{e_j}}}x}}{{\sigma _{i{e_j}}^2{2^{{R_s}}}{N_b}}}} )} ,
\end{split}
\end{equation}
where ${\cal {E}}_n$ represents the $n$-th non-empty subset of the elements of ${\cal {E}}$. Substituting (53) into (52) yields
\begin{equation}\label{equa54}
\begin{split}
P_{out,floor}^{sub} =& \sum\limits_{i = 1}^M {\sum\limits_{n = 1}^{{2^N} - 1} {\frac{{{{( - 1)}^{|{{\cal {E}}_n}| + 1}}}}{{\sigma _{ib}^2}}}}\int_0^\infty  {\exp ( - \frac{x}{{\sigma _{ib}^2}} )}\\
&\quad\quad\quad \times \exp (- \sum\limits_{{e_j} \in {{\cal {E}}_n}} {\frac{{{N_{{e_j}}}x}}{{\sigma _{i{e_j}}^2{2^{{R_s}}}{N_b}}}} )\\
&\quad\quad\quad \times \prod\limits_{\scriptstyle k \in {\cal {U}} \hfill \atop
\scriptstyle k \ne i \hfill} {\left( {1 - \exp ( - \frac{x}{{\sigma _{kb}^2}})} \right)} dx.
\end{split}
\end{equation}
Using the result of Appendix D, we have
\begin{equation}\label{equa55}
1 - \exp ( - \frac{x}{{\sigma _{kb}^2}}) \mathop  = \limits^1 \frac{x}{{\sigma _{kb}^2}},
\end{equation}
for $\lambda_{me} \to \infty  $, where $\mathop  = \limits^1 $ represents an equality with probability 1, and $x$ is a random variable with the following PDF
\begin{equation}\label{equa56}
g(x) = \frac{1}{{\sigma _{ib}^2}}\exp ( - \frac{x}{{\sigma _{ib}^2}} - \sum\limits_{{e_j} \in {{\cal {E}}_n}} {\frac{{{N_{{e_j}}}x}}{{\sigma _{i{e_j}}^2{2^{{R_s}}}{N_b}}}} ),
\end{equation}
wherein $0<x<\infty$. Hence, letting $\lambda_{me} \to \infty$ and substituting (55) into (54) yield (57)
\begin{figure*}
\begin{equation}\label{equa57}
\begin{split}
P_{out,floor}^{sub}
&= \sum\limits_{i = 1}^M {\sum\limits_{n = 1}^{{2^N} - 1} {\frac{{{{( - 1)}^{|{{\cal {E}}_n}| + 1}}}}{{{\theta _{ib}}}}\prod\limits_{\scriptstyle k \in {\cal {U}} \hfill \atop
  \scriptstyle k \ne i \hfill} {\frac{1}{{{\theta _{kb}}}}} {(\frac{1}{{{\theta _{ib}}{\lambda _{me}}}} + \sum\limits_{{e_j} \in {{\cal {E}}_n}} {\frac{{{N_{{e_j}}}}}{{{\theta _{i{e_j}}}{2^{{R_s}}}{N_b}}}} )^{ - M}}} }  \cdot {(\frac{1}{{{\lambda _{me}}}})^M} \\
 \end{split}
\end{equation}
\end{figure*}
for the uncoordinated case. Similarly to (31), the secrecy diversity order of the suboptimal user scheduling scheme is defined as
\begin{equation}\label{equa58}
{d_{sub}} =  - \mathop {\lim }\limits_{{\lambda _{me}} \to \infty } \frac{{\log (P_{out,floor}^{sub})}}{{\log ({\lambda _{me}})}}.
\end{equation}
Combining (57) and (58), we obtain the diversity order of the suboptimal user scheduling as
\begin{equation}\label{equa59}
{d_{sub}} = M,
\end{equation}
{for the uncoordinated case. Additionally, combining (3) and (34), we may obtain the lower and upper bounds on the secrecy outage probability floor of the suboptimal user scheduling scheme as
\begin{equation}\label{equa60}
P_{out,floor}^{lower}  \le P_{out,floor}^{sub}  = \mathop {\lim }\limits_{I \to \infty } P_{out}^{sub}  \le P_{out,floor}^{upper},
\end{equation}
where $P_{out}^{lower}$ and $P_{out}^{upper}$ are given by
\begin{equation}\label{equa61}
P_{out,floor}^{lower}  = \sum\limits_{i = 1}^M {\Pr \left(
\begin{split}
&{2^{R_s } \mathop {\max }\limits_{e_j  \in {\cal E}} \frac{{|h_{ie_j } |^2 }}{{\mathop {\max }\limits_{e_j  \in {\cal E}} N_{e_j } }} > \frac{1}{{N_b }}|h_{ib} |^2} ,\\
&\mathop {\max }\limits_{\scriptstyle k \in {\cal U} \hfill \atop
\scriptstyle k \ne i \hfill} |h_{kb} |^2  < |h_{ib} |^2
\end{split}
\right)},
\end{equation}
and
\begin{equation}\label{equa62}
P_{out,floor}^{upper}  = \sum\limits_{i = 1}^M {\Pr \left(
\begin{split}
&{2^{R_s } N\mathop {\max }\limits_{e_j  \in {\cal E}} \frac{{|h_{ie_j } |^2 }}{{\mathop {\min }\limits_{e_j  \in {\cal E}} N_{e_j } }} > \frac{1}{{N_b }}|h_{ib} |^2 },\\
&\mathop {\max }\limits_{\scriptstyle k \in {\cal U} \hfill \atop
\scriptstyle k \ne i \hfill} |h_{kb} |^2  < |h_{ib} |^2
\end{split}
\right)},
\end{equation}
for the coordinated case. Comparing (61) and (62) with (51) and using (57), we similarly obtain (63) and (64) at the top of the following page.
\begin{figure*}
\begin{equation}\label{equa63}
P_{out,floor}^{lower}  = \sum\limits_{i = 1}^M {\sum\limits_{n = 1}^{2^N  - 1} {\frac{{( - 1)^{|{\cal E}_n | + 1} }}{{\theta _{ib} }}\prod\limits_{\scriptstyle k \in {\cal U} \hfill \atop
  \scriptstyle k \ne i \hfill} {\frac{1}{{\theta _{kb} }}(\frac{1}{{\theta _{ib} \lambda _{me} }} + \sum\limits_{e_j  \in {\cal E}_n } {\frac{{\mathop {\max }\limits_{e_j  \in {\cal E}} N_{e_j } }}{{\theta _{ie_j } 2^{R_s } N_b }}} )^{ - M} } } }  \cdot (\frac{1}{{\lambda _{me} }})^M
\end{equation}
\end{figure*}

\begin{figure*}
\begin{equation}\label{equa64}
P_{out,floor}^{upper}  = \sum\limits_{i = 1}^M {\sum\limits_{n = 1}^{2^N  - 1} {\frac{{( - 1)^{|{\cal E}_n | + 1} }}{{\theta _{ib} }}\prod\limits_{\scriptstyle k \in {\cal U} \hfill \atop
  \scriptstyle k \ne i \hfill} {\frac{1}{{\theta _{kb} }}(\frac{1}{{\theta _{ib} \lambda _{me} }} + \sum\limits_{e_j  \in {\cal E}_n } {\frac{{\mathop {\min }\limits_{e_j  \in {\cal E}} N_{e_j } }}{{N\theta _{ie_j } 2^{R_s } N_b }}} )^{ - M} } } }  \cdot (\frac{1}{{\lambda _{me} }})^M
\end{equation}
\end{figure*}
Substituting (60) into (58) and using (63) and (64), we obtain the secrecy diversity of the suboptimal user scheduling as
\begin{equation}\nonumber
M \le d_{sub}  \le M,
\end{equation}
which results in
\begin{equation}\label{equa65}
 d_{sub}  = M,
\end{equation}
for the coordinated case. It can be observed from (59) and (65) that no matter whether the eavesdroppers are coordinated or not, the suboptimal user scheduling scheme achieves the diversity order of $M$, which is the same as the optimal user scheduling approach. It is worth mentioning that the suboptimal user scheduling only needs the CSIs of CUs-CBS links. However, the optimal user scheduling assumes that the CSIs of all links from CUs to CBS, to PR and to ${\textrm{E}}_j$ are known, which makes it challenging to be applied in practical cognitive radio systems. Therefore, from a practical perspective, the suboptimal scheduling scheme is more attractive than the optimal scheduling.}

\section{Numerical Results and Discussions}
In this section, we present numerical comparison among the round-robin scheduling, the optimal user scheduling and the suboptimal user scheduling in terms of secrecy outage probability. Throughout the numerical secrecy outage evaluation, we assume that the background noise and interference received at any node in the cognitive radio network shown in Fig. 1 (including CBS and $N$ eavesdroppers) have the same variance, i.e., $N_b=N_{e_j}$ for $e_j \in {\cal {E}}$. For notational convenience, let $\lambda_I$ denote the ratio of the maximum allowable interference power $I$ to the noise variance $N_b$, i.e., $\gamma_{I}=I/N_b$.

\begin{figure}
  \centering
  {\includegraphics[scale=0.58]{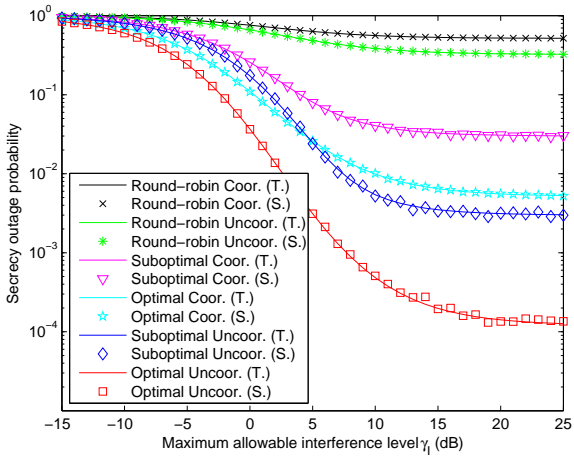}\\
  \caption{{Secrecy outage probability versus maximum allowable interference level $\gamma_{I}$ of the round-robin scheduling, the suboptimal user scheduling and the optimal user scheduling schemes for both the uncoordinated and coordinated cases with $M=8$, $N=4$, $R_s=1{\textrm{ bit/s/Hz}}$, $\lambda_{me}=10{\textrm{ dB}}$, and $\sigma^2_{ib}=\sigma^2_{ip}=\theta_{ib}=\theta_{ie_j}=1$.}}\label{Fig2}}
\end{figure}
Fig. 2 shows the secrecy outage probability versus the maximum allowable interference level $\gamma_{I}$ of the round-robin scheduling as well as the optimal and suboptimal scheduling schemes {for the uncoordinated and coordinated cases by using (13), (18), (26) and (28)}. Simulation results of the secrecy outage probability for these three schemes are also provided in this figure. It is observed from Fig. 2 that as the maximum allowable interference level $\gamma_I$ increases, the secrecy outage probabilities of the round-robin scheduling, the suboptimal user scheduling and the optimal user scheduling schemes all decrease. This can be explained that with an increasing $\gamma_I$, CUs are allowed to transmit with higher power, leading to a decrease of the secrecy outage probability. One can see from Fig. 2 that as $\gamma_I$ increases beyond a certain value, these three schemes converge to their respective secrecy outage probability floors, where the optimal and suboptimal scheduling schemes both have a lower secrecy outage floor than the round-robin scheduling. Moreover, {for both the uncoordinated and coordinated cases}, the optimal user scheduling strictly outperforms the suboptimal user scheduling in terms of the secrecy outage probability. {Fig. 2 also illustrates that the secrecy outage performance of the round-robin scheduling as well as the optimal and suboptimal scheduling corresponding to the uncoordinated eavesdroppers is expectedly better than that of these three schemes corresponding to the coordinated eavesdroppers. This means that the eavesdroppers may collaborate with each other for the sake of degrading the secrecy outage performance of cognitive transmissions. }In addition, the simulation results match well the theoretical secrecy outage probabilities, confirming the correctness of the secrecy outage analysis.

\begin{figure}
  \centering
  {\includegraphics[scale=0.58]{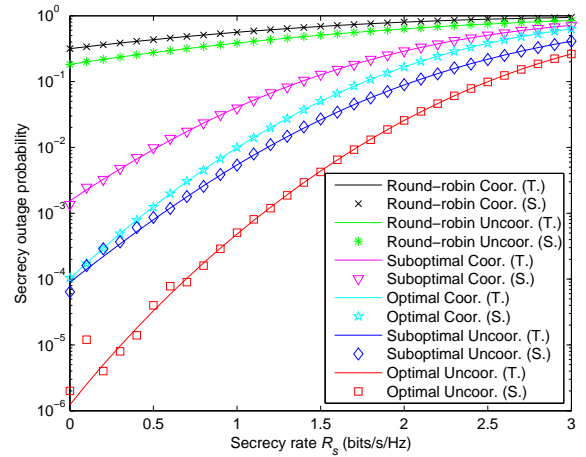}\\
  \caption{{Secrecy outage probability versus secrecy rate $R_s$ of the round-robin scheduling, the suboptimal user scheduling and the optimal user scheduling schemes for both the uncoordinated and coordinated cases with $M=8$, $N=4$, $\lambda_{me}=10{\textrm{ dB}}$, $\gamma_I=10{\textrm{ dB}}$, and $\sigma^2_{ib}=\sigma^2_{ip}=\theta_{ib}=\theta_{ie_j}=1$.}}\label{Fig3}}
\end{figure}
Fig. 3 depicts the secrecy outage probability versus secrecy rate $R_s$ of the round-robin scheduling, the suboptimal user scheduling and the optimal user scheduling schemes {for both the uncoordinated and coordinated cases}. As shown in Fig. 3, with an increasing secrecy rate $R_s$, the secrecy outage probabilities of these three schemes {in the presence of coordinated or uncoordinated eavesdroppers} all increase accordingly. In other words, when a higher secrecy rate $R_s$ is adopted by CUs for better throughput performance, it is less likely to achieve the perfect secure transmission against eavesdropping attacks. {It is also seen from Fig. 3 that for both the coordinated and uncoordinated cases, the optimal user scheduling scheme achieves the best secrecy outage performance and the round-robin scheduling performs the worst across the whole secrecy rate region.}

\begin{figure}
  \centering
  {\includegraphics[scale=0.58]{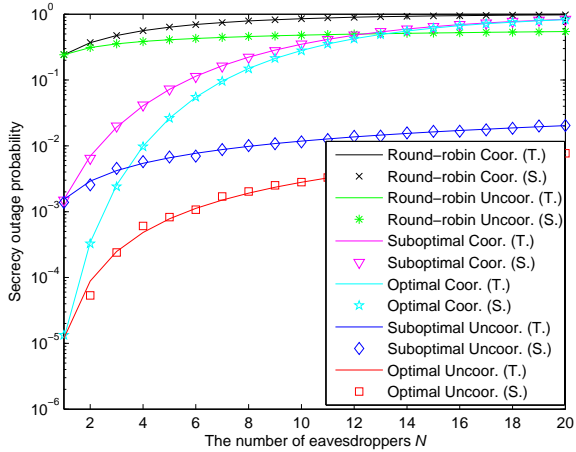}\\
  \caption{{Secrecy outage probability versus the number of eavesdroppers $N$ of the round-robin scheduling, the suboptimal user scheduling and the optimal user scheduling schemes for both the uncoordinated and coordinated cases with $M=8$, $R_s=1{\textrm{ bit/s/Hz}}$, $\lambda_{me}=10{\textrm{ dB}}$, $\gamma_I=10{\textrm{ dB}}$, and $\sigma^2_{ib}=\sigma^2_{ip}=\theta_{ib}=\theta_{ie_j}=1$.}}\label{Fig4}}
\end{figure}
In Fig. 4, we show the secrecy outage probability versus the number of eavesdroppers $N$ of the round-robin scheduling, the suboptimal user scheduling and the optimal user scheduling schemes {for both the uncoordinated and coordinated eavesdroppers}. One can observe from Fig. 4 that as the number of eavesdroppers $N$ increases, the secrecy outage probabilities of these three schemes all increase {for both the uncoordinated and coordinated cases}. Nevertheless, given a certain number of {uncoordinated (or coordinated) eavesdroppers}, the optimal and suboptimal user scheduling schemes both perform better than the round-robin scheduling in terms of the secrecy outage probability.

\begin{figure}
  \centering
  {\includegraphics[scale=0.58]{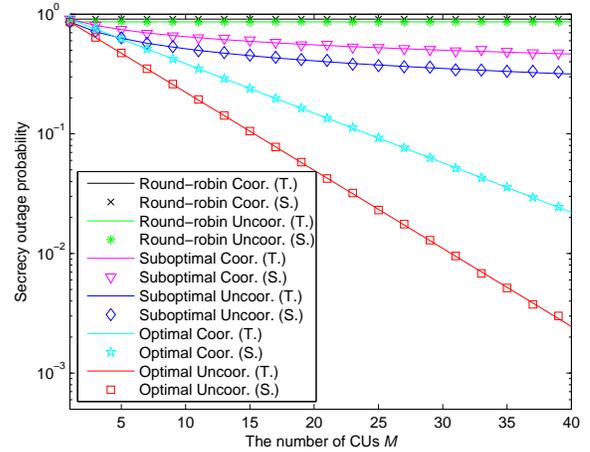}\\
  \caption{{Secrecy outage probability versus the number of CUs $M$ of the round-robin scheduling, the suboptimal user scheduling and the optimal user scheduling schemes for both the uncoordinated and coordinated cases with $N=2$, $R_s=0.2{\textrm{ bit/s/Hz}}$, $\lambda_{me}=-3{\textrm{ dB}}$, $\gamma_I=10{\textrm{ dB}}$, and $\sigma^2_{ib}=\sigma^2_{ip}=\theta_{ib}=\theta_{ie_j}=1$.}}\label{Fig5}}
\end{figure}
Fig. 5 illustrates the secrecy outage probability versus the number of CUs $M$ of the round-robin scheduling as well as the optimal and suboptimal scheduling schemes {for both the uncoordinated and coordinated cases with a low MER of $\lambda_{me}=-3{\textrm{ dB}}$, which means that the average channel gain of the eavesdroppers is two times better than that of the legitimate CUs. It is shown from Fig. 5 that upon increasing the number of CUs, the secrecy outage probability of the round-robin scheduling scheme keeps unchanged for both the uncoordinated and coordinated cases, showing no security benefits achieved with an increasing number of CUs.} By contrast, the secrecy outage probabilities of the optimal and suboptimal user scheduling schemes both significantly decrease, as the number of CUs increases. {Therefore, when either the optimal or suboptimal user scheduling is adopted, the security of cognitive transmissions may be improved by increasing the number of CUs. In other words, upon increasing the number of CUs, any target secrecy outage probability can be guaranteed by relying on the optimal or suboptimal user scheduling scheme even with a very low MER.}


\section{Conclusion}
In this paper, we have studied the secrecy outage and diversity performance of a multi-user multi-eavesdropper cognitive radio system, where CUs transmit to a common CBS with a primary QoS constraint {in the presence of multiple coordinated or uncoordinated eavesdroppers}. We have presented the round-robin scheduling, optimal user scheduling and suboptimal user scheduling schemes to protect the CUs-CBS transmissions against the eavesdropping attacks. Closed-form expressions of the secrecy outage probability of these three schemes have been derived {for both the coordinated and uncoordinated cases}. We have also conducted the secrecy diversity analysis of the round-robin scheduling as well as the optimal and suboptimal user scheduling in the presence of {the coordinated and uncoordinated eavesdroppers}. It has been proven that {no matter whether the eavesdroppers collaborate or not,} the round-robin scheduling achieves the secrecy diversity order of only one, whereas the optimal and suboptimal scheduling schemes both obtain the diversity order of $M$, where $M$ is the number of CUs. Numerical results have demonstrated that as the maximum allowable interference level increases, the secrecy outage performance of the round-robin scheduling as well as the optimal and suboptimal user scheduling improves accordingly. Additionally, {for both the coordinated and uncoordinated cases, the optimal and suboptimal user scheduling schemes significantly outperform the round-robin scheduling approach in terms of the secrecy outage probability.}

{In the present paper, we only studied the physical-layer security of CUs-CBS transmissions without considering the PUs' security for the primary network. In cognitive radio networks, CUs and PUs typically share the same spectrum band, thus the eavesdroppers may tap both the CUs' and PUs' transmissions. It is thus of high interest to investigate the impact of the uncoordinated and coordinated eavesdroppers on the security of both CUs and PUs. Considering the fact that PUs have a higher priority than CUs in accessing the licensed spectrum, we may impose a security constraint on the PUs' transmissions (e.g., the secrecy capacity, secrecy outage probability, etc.) with an objective of maximizing the CUs' security performance with the aid of some signal processing techniques, e.g. multi-user scheduling and beamforming. Besides, due to estimation errors in practical channel estimators, the perfect CSI is impossible to be obtained. It is interesting to further examine the impact of CSI estimation errors on the secrecy outage performance of various scheduling schemes. We leave these interesting problems for our future work.}

\appendices
\section{Proof of (11) and (12)}
For notational convenience, let $X$ denote term $\frac{1}{{{2^{{R_s}}}{N_b}}}|{h_{ib}}{|^2} - \frac{{{2^{{R_s}}} - 1}}{{{2^{{R_s}}}I}}|{h_{ip}}{|^2}$, i.e., $X = \frac{1}{{{2^{{R_s}}}{N_b}}}|{h_{ib}}{|^2} - \frac{{{2^{{R_s}}} - 1}}{{{2^{{R_s}}}I}}|{h_{ip}}{|^2} $. Note that random variables $|h_{ib}|^2$ and $|h_{ip}|^2$ are exponentially distributed and independent of each other. Denoting ${X_1} = |{h_{ib}}{|^2}$ and ${X_2} = |{h_{ip}}{|^2}$, we obtain the cumulative distribution function (CDF) of $X$ as
\begin{equation}
\Pr \left( {X < x} \right) = \Pr \left( {\frac{1}{{{2^{{R_s}}}{N_b}}}{X_1} - \frac{{{2^{{R_s}}} - 1}}{{{2^{{R_s}}}I}}{X_2} < x} \right),\tag{A.1}\label{A.1}
\end{equation}
where $ - \infty  < x < \infty $. For $x < 0$, (A.1) can be given by
\begin{equation}
\begin{split}
\Pr \left( {X < x} \right)
&= \frac{{\sigma _{ip}^2({2^{{R_s}}} - 1){N_b}}}{{\sigma _{ip}^2({2^{{R_s}}} - 1){N_b} + \sigma _{ib}^2I}}\\
&\quad\times\exp \left( {\frac{{{2^{{R_s}}}I}}{{({2^{{R_s}}} - 1)\sigma _{ip}^2}}x} \right),
\end{split}\tag{A.2}\label{A.2}
\end{equation}
where $\sigma _{ib}^2 = E(|{h_{ib}}{|^2})$ and $\sigma _{ip}^2 = E(|{h_{ip}}{|^2})$. Besides, for $x > 0$, (A.1) is obtained as
\begin{equation}
\begin{split}
\Pr \left( {X < x} \right)
& = 1 - \frac{{\sigma _{ib}^2I}}{{\sigma _{ib}^2I + \sigma _{ip}^2({2^{{R_s}}} - 1){N_b}}}\\
&\quad\quad\quad\times\exp \left( { - \frac{{{2^{{R_s}}}{N_b}}}{{\sigma _{ib}^2}}x} \right).
\end{split}\tag{A.3}\label{A.3}
\end{equation}
Combining (A.2) and (A.3), we can prove that the CDF of $X$ is first-order differentiable for $-\infty < x < \infty$ and obtain the probability density function (PDF) of $X$ as
\begin{equation}
{f_X}(x) =  \begin{cases}
 \dfrac{{{2^{{R_s}}}{N_b}I}}{A}\exp \left( {\dfrac{{{2^{{R_s}}}Ix}}{{\sigma _{ip}^2({2^{{R_s}}} - 1)}}} \right),&x<0 \vspace{2 mm} \\
 \dfrac{{{2^{{R_s}}}{N_b}I}}{A}\exp \left( { - \dfrac{{{2^{{R_s}}}{N_b}}}{{\sigma _{ib}^2}}x} \right),&x>0 , \\
 \end{cases} \tag{A.4}\label{A.4}
\end{equation}
where $A={\sigma _{ip}^2({2^{{R_s}}} - 1){N_b} + \sigma _{ib}^2I}$. Noting that ${|{h_{i{e_j}}}{|^2}}$ for $e_j \in {\cal{E}}$ are independent exponentially distributed random variables with respective means of ${\sigma _{i{e_j}}^2}$ for different eavesdroppers, we can rewrite (10) as
\begin{equation}
\begin{split}
{P_{out,i}} &= 1 - \Pr \left( {\mathop {\max }\limits_{{e_j} \in {\cal{E}}} \frac{{|{h_{i{e_j}}}{|^2}}}{{{N_{{e_j}}}}} < X} \right) \\
&= 1 - \int_0^\infty  {\prod\limits_{{e_j} \in {\cal {E}}} {\left( {1 - \exp ( - \frac{{N_{{e_j}}}x}{{\sigma _{i{e_j}}^2}})} \right)} {f_X}(x)dx}.
 \end{split} \tag{A.5}\label{A.5}
\end{equation}
Substituting (A.4) into (A.5) yields
\begin{equation}
\begin{split}
{P_{out,i}} = &1 - \frac{{{2^{{R_s}}}{N_b}I}}{{\sigma _{ib}^2I + \sigma _{ip}^2({2^{{R_s}}} - 1){N_b}}}\\
&\quad\times\int_0^\infty  {\prod\limits_{{e_j} \in {\cal {E}}} {\left( {1 - \exp ( - \frac{{N_{{e_j}}}x}{{\sigma _{i{e_j}}^2}})} \right)} }\\
&\quad\quad\quad\times\exp ( { - \frac{{{2^{{R_s}}}{N_b}}}{{\sigma _{ib}^2}}x} )dx,
\end{split}\tag{A.6}\label{A.6}
\end{equation}
where term $\prod\limits_{{e_j} \in {\cal {E}}} {\left( {1 - \exp ( - \frac{{N_{{e_j}}}x}{{\sigma _{i{e_j}}^2}})} \right)} $ can be expanded with the binomial theorem as
\begin{equation}
\begin{split}
&\prod\limits_{{e_j} \in {\cal {E}}} {\left( {1 - \exp ( - \frac{{N_{{e_j}}}x}{{\sigma _{i{e_j}}^2}})} \right)}  \\
&= 1 - \sum\limits_{n = 1}^{{2^N} - 1} {{{( - 1)}^{|{{\cal {E}}_n}|+1}}\exp ( - \sum\limits_{{e_j} \in {{\cal {E}}_n}} {\frac{{N_{{e_j}}}x}{{\sigma _{i{e_j}}^2}}} )},
\end{split}\tag{A.7}\label{A.7}
\end{equation}
where $N$ is the number of eavesdroppers, ${\cal {E}}_n$ represents the $n$-th non-empty subset of the elements of ${\cal {E}}$, and $|{\cal {E}}_n|$ is the cardinality of set ${\cal {E}}_n$. Substituting (A.7) into (A.6) and performing the integration yield
\begin{equation}
\begin{split}
{P_{out,i}} = \frac{{\sigma _{ip}^2({2^{{R_s}}} - 1){N_b} + \sum\limits_{n = 1}^{{2^N} - 1} {\frac{{{{( - 1)}^{|{{\cal {E}}_n}| + 1}}{2^{{R_s}}}{N_b}I}}{{\sigma _{ib}^{ - 2}{2^{{R_s}}}{N_b} + \sum\limits_{{e_j} \in {{\cal {E}}_n}} {{{(\sigma _{i{e_j}}^{-2}{N_{{e_j}}})}}} }}} }}{{\sigma _{ib}^2I + \sigma _{ip}^2({2^{{R_s}}} - 1){N_b}}},\\
 \end{split}\tag{A.8}\label{A.8}
\end{equation}
which is (11). {Additionally, substituting (3), (7) and (8) into (9) and assuming different eavesdroppers with the same noise variance of $N_{e_j}=N_e$, we have
\begin{equation}
P_{out,i}  = \Pr \left( {\sum\limits_{e_j  \in {\cal E}} {\frac{{|h_{ie_j } |^2 }}{{N_e }}}  > \frac{1}{{2^{R_s } N_b }}|h_{ib} |^2  - \frac{{2^{R_s }  - 1}}{{2^{R_s } I}}|h_{ip} |^2 } \right)
,\tag{A.9}\label{A.9}
\end{equation}
for the coordinated case. For notational convenience, we denote $X = \frac{1}{{2^{R_s } N_b }}|h_{ib} |^2  - \frac{{2^{R_s }  - 1}}{{2^{R_s } I}}|h_{ip} |^2$ and $Y = \sum\limits_{e_j  \in {\cal E}} {\frac{{|h_{ie_j } |^2 }}{{N_e }}}$, where the PDF of $X$ is given by (A.4). Moreover, considering that random variables $|h_{ie_j}|^2$ for $e_j \in {\cal E}$ are i.i.d. with the same mean of $\sigma^2_{ie}$, we obtain that $Y$ is Gamma distributed with the mean of $\frac{{N\sigma _{ie}^2 }}{{N_e }}$, whose PDF is given by
\begin{equation}
f_Y (y) = \frac{{(N_e )^N }}{{\Gamma (N)\sigma _{ie}^{2N} }}y^{N - 1} \exp ( - \frac{{N_e y}}{{\sigma _{ie}^2 }})
,\tag{A.10}\label{A.10}
\end{equation}
for $y>0$, where $N$ is the number of eavesdroppers. Noting that random variables $X$ and $Y$ are independent and combining (A.4), (A.9) and (A.10), we obtain
\begin{equation}
\begin{split}
&P_{out,i}  = 1 - \Pr \left( {Y < X} \right) \\
&=  1 - \frac{{\sigma _{ib}^2 I}}{{\sigma _{ib}^2 I + \sigma _{ip}^2 (2^{R_s}  - 1)N_b }}(1 + \frac{{2^{R_s} \sigma _{ie}^2 N_b }}{{\sigma _{ib}^2 N_e }})^{ - N},
 \end{split}\tag{A.11}\label{A.11}
\end{equation}
which is (12).}

\section{Derivation of (26)}
Considering that $|{h_{ib}}{|^2}$ and $|{h_{ip}}{|^2}$ are independent exponentially distributed and denoting $X=|{h_{ib}}{|^2}$ and $Y=|{h_{ip}}{|^2}$, we can easily obtain the joint PDF of random variables $(X,{\textrm{ }}Y)$ as
\begin{equation}
f(x,y) = \frac{1}{{\sigma _{ib}^2\sigma _{ip}^2}}\exp ( - \frac{x}{{\sigma _{ib}^2}} - \frac{y}{{\sigma _{ip}^2}}),\tag{B.1}\label{B.1}
\end{equation}
for $(x>0,{\textrm{ }}y>0)$. Thus, we can rewrite (25) as
\begin{equation}
\begin{split}
&P_{out}^{sub} = \sum\limits_{i = 1}^M {\Pr \left(
\begin{split}
&{{2^{{R_s}}}I\mathop {\max }\limits_{{e_j} \in {\cal{E}}} \frac{{|{h_{i{e_j}}}{|^2}}}{{{N_{{e_j}}}}} > \frac{I}{{{N_b}}}X - ({2^{{R_s}}} - 1)Y},\\
&{\textrm{ }}\mathop {\max }\limits_{\scriptstyle k \in {\cal{U}} \hfill \atop\scriptstyle k \ne i \hfill} |{h_{kb}}{|^2} < X
\end{split}
\right)}  \\
&= \sum\limits_{i = 1}^M {\iint\limits {{\Pr \left( {{2^{{R_s}}}I\mathop {\max }\limits_{{e_j} \in {\cal{E}}} \frac{{|{h_{i{e_j}}}{|^2}}}{{{N_{{e_j}}}}} > \frac{I}{{{N_b}}}x - ({2^{{R_s}}} - 1)y} \right)}}}\\
&\quad\quad\quad\quad\quad \times {{{\Pr \left( {\mathop {\max }\limits_{\scriptstyle k \in {\cal{U}} \hfill \atop
  \scriptstyle k \ne i \hfill} |{h_{kb}}{|^2} < x} \right)f(x,y)dxdy}} },  \\
\end{split}\tag{B.2}\label{B.2}
\end{equation}
where the second equation arises from the fact that random variables $|{h_{i{e_j}}}{|^2}$ and $|{h_{kb}}{|^2}$ (for ${e_j} \in {\cal E}$ and $k \in {\cal U}$) are independent of each other. Noting that $|{h_{i{e_j}}}{|^2}$ and $|{h_{kb}}{|^2}$ are exponential random variables with respective means $\sigma _{i{e_j}}^2$ and $\sigma _{kb}^2$, (B.2) can be further obtained as (B.3) at the top of the following page,
\begin{figure*}
\begin{equation}
\begin{split}
 P_{out}^{sub} &= \sum\limits_{i = 1}^M {\iint\limits_{{\Omega}}{{\left[ {1 - \prod\limits_{{e_j} \in {\cal{E}}} {\left( {1 - \exp ( - \frac{{{N_{{e_j}}}N_b^{ - 1}Ix - {N_{{e_j}}}({2^{{R_s}}} - 1)y}}{{\sigma _{i{e_j}}^2{2^{{R_s}}}I}})} \right)} } \right]}}} \prod\limits_{\scriptstyle k \in {\cal{U}} \hfill \atop
  \scriptstyle k \ne i \hfill} {\left( {1 - \exp ( - \frac{x}{{\sigma _{kb}^2}})} \right)} f(x,y)dxdy\\
&\quad +\sum\limits_{i = 1}^M {\iint\limits_\Phi  {{\prod\limits_{\scriptstyle k \in {\cal {U}} \hfill \atop
  \scriptstyle k \ne i \hfill} {\left( {1 - \exp ( - \frac{x}{{\sigma _{kb}^2}})} \right)} f(x,y)dxdy}} }
\end{split}\tag{B.3}\label{B.3}
\end{equation}
\end{figure*}
where $\Omega  = \left\{ {(x,y)|N_b^{ - 1}Ix - ({2^{{R_s}}} - 1)y > 0} \right\}$ and $\Phi  = \left\{ {(x,y)|N_b^{ - 1}Ix - ({2^{{R_s}}} - 1)y < 0} \right\}$. By using the binomial theorem, term $\prod\limits_{{e_j} \in {\cal{E}}} {\left( {1 - \exp ( - \frac{{{N_{{e_j}}}N_b^{ - 1}Ix - {N_{{e_j}}}({2^{{R_s}}} - 1)y}}{{\sigma _{i{e_j}}^2{2^{{R_s}}}I}})} \right)}$ is expanded by
\begin{equation}
\begin{split}
&\prod\limits_{{e_j} \in {\cal{E}}} {\left( {1 - \exp ( - \frac{{{N_{{e_j}}}N_b^{ - 1}Ix - {N_{{e_j}}}({2^{{R_s}}} - 1)y}}{{\sigma _{i{e_j}}^2{2^{{R_s}}}I}})} \right)}   = 1 -\\
&\sum\limits_{n = 1}^{{2^N} - 1} {{{( - 1)}^{|{{\cal{E}}_n}| + 1}}\exp ( - \sum\limits_{{e_j} \in {\cal{E}}_n} {\frac{{{N_{{e_j}}}N_b^{ - 1}Ix - {N_{{e_j}}}({2^{{R_s}}} - 1)y}}{{\sigma _{i{e_j}}^2{2^{{R_s}}}I}}} )}, \end{split}\tag{B.4}\label{B.4}
\end{equation}
where ${\cal {E}}_n$ represents the $n$-th non-empty subset of the elements of ${\cal {E}}$ and $|{\cal {E}}_n|$ is the cardinality of set ${\cal {E}}_n$. Similarly, we can obtain term $\prod\limits_{\scriptstyle k \in {\cal{U}} \hfill \atop
  \scriptstyle k \ne i \hfill} {\left( {1 - \exp ( - \frac{x}{{\sigma _{kb}^2}})} \right)} $ as
\begin{equation}
\begin{split}
&\prod\limits_{\scriptstyle k \in {\cal{U}} \hfill \atop\scriptstyle k \ne i \hfill} {\left( {1 - \exp ( - \frac{x}{{\sigma _{kb}^2}})} \right)}  \\
&= 1 - \sum\limits_{m = 1}^{{2^{M - 1}} - 1} {{{( - 1)}^{|{{\cal{U}}_m}| + 1}}\exp ( - \sum\limits_{k \in {{\cal{U}}_m}} {\frac{x}{{\sigma _{kb}^2}}} )},
\end{split}\tag{B.5}\label{B.5}
\end{equation}
where ${\cal {U}}_m$ represents the $m$-th non-empty subset of the elements of ${\cal {U}}-\{{{\textrm{CU}}_i}\}$, `$-$' represents the set difference, and $|{\cal {U}}_m|$ is the cardinality of set ${\cal {U}}_m$. Substituting (B.4) and (B.5) into (B.3) yields
\begin{equation}
\begin{split}
P_{out}^{sub} = &\sum\limits_{i = 1}^M {\sum\limits_{n = 1}^{{2^N} - 1} {{{( - 1)}^{|{{\cal{E}}_n}| + 1}}\left( {P_{out,I}^{sub} - P_{out,II}^{sub}} \right)} } \\
&+ \sum\limits_{i = 1}^M {P_{out,III}^{sub}} ,
\end{split}
\tag{B.6}\label{B.6}
\end{equation}
where $P_{out,I}^{sub}$, $P_{out,II}^{sub}$ and ${P_{out,III}^{sub}}$ are given by
\begin{equation}
\begin{split}
P_{out,I}^{sub} =&\iint\limits_{\Omega} {{\exp ( - \sum\limits_{{e_j} \in {{\cal{E}}_n}} {\frac{{{N_{{e_j}}}N_b^{ - 1}Ix - {N_{{e_j}}}({2^{{R_s}}} - 1)y}}{{\sigma _{i{e_j}}^2{2^{{R_s}}}I}}} )}}\\
&\quad\quad\times f(x,y)dxdy,
\end{split}
\tag{B.7}\label{B.7}
\end{equation}
and
\begin{equation}
\begin{split}
P_{out,II}^{sub} =& \sum\limits_{m = 1}^{{2^{M - 1}} - 1} {{{( - 1)}^{|{{\cal{U}}_m}| + 1}}\iint\limits_{\Omega} {{\exp ( - \sum\limits_{k \in {{\cal{U}}_m}} {\frac{x}{{\sigma _{kb}^2}}}  )}} }\\
& \quad\quad \times \exp ( - \sum\limits_{{e_j} \in {{\cal{E}}_n}} {\frac{{{N_{{e_j}}}N_b^{ - 1}Ix - {N_{{e_j}}}({2^{{R_s}}} - 1)y}}{{\sigma _{i{e_j}}^2{2^{{R_s}}}I}}} ) \\
&\quad\quad\times f(x,y)dxdy, \\
\end{split}\tag{B.8}\label{B.8}
\end{equation}
and
\begin{equation}
\begin{split}
P_{out,III}^{sub} =&\iint\limits_\Phi  {\left[ {1 - \sum\limits_{m = 1}^{{2^{M - 1}} - 1} {{{( - 1)}^{|{{\cal {U}}_m}|{\rm{ + }}1}}\exp ( - \sum\limits_{k \in {{\cal {U}}_m}} {\frac{x}{{\sigma _{kb}^2}}} )} } \right]}\\
&\quad\quad\times f(x,y)dxdy.
\end{split}
\tag{B.9}\label{B.9}
\end{equation}
Combining (B.1) and (B.7), we obtain
\begin{equation}
\begin{split}
 P_{out,I}^{sub} = &\int_0^\infty  {\frac{1}{{\sigma _{ib}^2}}\exp ( - \frac{x}{{\sigma _{ib}^2}} - \sum\limits_{{e_j} \in {{\cal{E}}_n}} {\frac{{{N_{{e_j}}}x}}{{\sigma _{i{e_j}}^2{2^{{R_s}}}{N_b}}}} )dx}  \\
&\quad \times \int_0^{\frac{{Ix}}{{({2^{{R_s}}} - 1){N_b}}}} {\frac{1}{{\sigma _{ip}^2}}\exp ( - \frac{y}{{\sigma _{ip}^2}})}\\
&\quad\quad\quad\times \exp (\sum\limits_{{e_j} \in {{\cal{E}}_n}} {\frac{{{N_{{e_j}}}({2^{{R_s}}} - 1)y}}{{\sigma _{i{e_j}}^2{2^{{R_s}}}I}}} )dy,  \\
\end{split}\tag{B.10}\label{B.10}
\end{equation}
which is further computed as
\begin{equation}
\begin{split}
 P_{out,I}^{sub}
=\frac{{I{(\frac{1}{{\sigma _{ib}^2}} + \sum\limits_{{e_j} \in {E_n}} {\frac{{{N_{{e_j}}}}}{{\sigma _{i{e_j}}^2{2^{{R_s}}}{N_b}}}} )^{ - 2}}}}{{\sigma _{ib}^2\sigma _{ip}^2({2^{{R_s}}} - 1){N_b}}}, \\
\end{split}\tag{B.11}\label{B.11}
\end{equation}
for $\sum\limits_{{e_j} \in {{\cal{E}}_n}} {\frac{{({2^{{R_s}}} - 1){N_{{e_j}}}}}{{\sigma _{i{e_j}}^2{2^{{R_s}}}I}}}  = \frac{1}{{\sigma _{ip}^2}}$. Moreover, if $\sum\limits_{{e_j} \in {{\cal{E}}_n}} {\frac{{({2^{{R_s}}} - 1){N_{{e_j}}}}}{{\sigma _{i{e_j}}^2{2^{{R_s}}}I}}}  \ne \frac{1}{{\sigma _{ip}^2}}$, we can obtain $ P_{out,I}^{sub} $ from (B.10) as
\begin{equation}
P_{out,I}^{sub} = \frac{{{(1 + \sum\limits_{{e_j} \in {{\cal{E}}_n}} {\frac{{\sigma _{ib}^2{N_{{e_j}}}}}{{\sigma _{i{e_j}}^2{2^{{R_s}}}{N_b}}}} )^{ - 1}} - {(1 + \frac{{\sigma _{ib}^2I}}{{\sigma _{ip}^2({2^{{R_s}}} - 1){N_b}}})^{ - 1}}}}{{1 - \sum\limits_{{e_j} \in {{\cal{E}}_n}} {\frac{{\sigma _{ip}^2{N_{{e_j}}}({2^{{R_s}}} - 1)}}{{\sigma _{i{e_j}}^2{2^{{R_s}}}I}}} }}.\tag{B.12}\label{B.12}
\end{equation}
Similarly, substituting (B.1) into (B.8) yields
\begin{equation}
P_{out,II}^{sub} = \sum\limits_{m = 1}^{{2^{M - 1}} - 1} {\frac{{{{( - 1)}^{|{{\cal{U}}_m}| + 1}}I{(B + \sum\limits_{{e_j} \in {{\cal{E}}_n}} {\frac{{{N_{{e_j}}}}}{{\sigma _{i{e_j}}^2{2^{{R_s}}}{N_b}}}} )^{ - 2}}}}{{\sigma _{ib}^2\sigma _{ip}^2({2^{{R_s}}} - 1){N_b}}}},\tag{B.13}\label{B.13}
\end{equation}
for $\sum\limits_{{e_j} \in {{\cal{{\cal{E}}}}_n}} {\frac{{({2^{{R_s}}} - 1){N_{{e_j}}}}}{{\sigma _{i{e_j}}^2{2^{{R_s}}}I}}}  = \frac{1}{{\sigma _{ip}^2}}$, wherein $B=\frac{1}{{\sigma _{ib}^2}} + \sum\limits_{k \in {{\cal{U}}_m}} {\frac{1}{{\sigma _{kb}^2}}}$. In case of $\sum\limits_{{e_j} \in {{\cal{{\cal{E}}}}_n}} {\frac{{({2^{{R_s}}} - 1){N_{{e_j}}}}}{{\sigma _{i{e_j}}^2{2^{{R_s}}}I}}} \ne \frac{1}{{\sigma _{ip}^2}}$, we obtain $P_{out,II}^{sub}$ as
\begin{equation}
\begin{split}
P_{out,II}^{sub} &= \sum\limits_{m = 1}^{{2^{M - 1}} - 1} {{{( - 1)}^{|{{\cal{U}}_m}| + 1}}\frac{{{(\sigma^2_{ib}B  + \sum\limits_{{e_j} \in {{\cal{E}}_n}} {\frac{{\sigma _{ib}^2{N_{{e_j}}}}}{{\sigma _{i{e_j}}^2{2^{{R_s}}}{N_b}}}} )^{ - 1}}}}{{1 - \sum\limits_{{e_j} \in {{\cal{E}}_n}} {\frac{{\sigma _{ip}^2{N_{{e_j}}}({2^{{R_s}}} - 1)}}{{\sigma _{i{e_j}}^2{2^{{R_s}}}I}}} }}}  \\
&- \sum\limits_{m = 1}^{{2^{M - 1}} - 1} {{{( - 1)}^{|{{\cal{U}}_m}| + 1}}\frac{{{(\sigma^2_{ib}B + \frac{{\sigma _{ib}^2I}}{{\sigma _{ip}^2({2^{{R_s}}} - 1){N_b}}})^{ - 1}}}}{{1 - \sum\limits_{{e_j} \in {{\cal{E}}_n}} {\frac{{\sigma _{ip}^2{N_{{e_j}}}({2^{{R_s}}} - 1)}}{{\sigma _{i{e_j}}^2{2^{{R_s}}}I}}} }}} . \\
\end{split}\tag{B.14}\label{B.14}
\end{equation}
In addition, combining (B.1) and (B.9), we obtain $P_{out,III}^{sub}$ as
\begin{equation}
\begin{split}
&P_{out,III}^{sub}
= {\left( {1 + \frac{{\sigma _{ib}^2I}}{{\sigma _{ip}^2({2^{{R_s}}} - 1){N_b}}}} \right)^{ - 1}} \\
&- {\sum\limits_{m = 1}^{{2^{M - 1}} - 1} {{{( - 1)}^{|{{\cal {U}}_m}|{\rm{ + }}1}}\left( {\sigma^2_{ib}B  + \frac{{\sigma _{ib}^2I}}{{\sigma _{ip}^2({2^{{R_s}}} - 1){N_b}}}} \right)} ^{ - 1}}, \\
\end{split}\tag{B.15}\label{B.15}
\end{equation}
where $B=\frac{1}{{\sigma _{ib}^2}} + \sum\limits_{k \in {{\cal{U}}_m}} {\frac{1}{{\sigma _{kb}^2}}}$. Finally, substituting (B.11)-(B.15) into (B.6) yields (26).

{\section{Derivation of (28)}
Denoting $|{h_{ib}}{|^2}=X$, $|{h_{ip}}{|^2}=Y$ and $\sum\limits_{e_j  \in {\cal E}} {|h_{ie_j } |^2 }=Z$, we can rewrite (27) as
\begin{equation}
P_{out}^{sub}  = \sum\limits_{i = 1}^M {\Pr \left(
\begin{split}
&{Z > \frac{{N_e }}{{2^{R_s } N_b }}X - \frac{{(2^{R_s }  - 1)N_e }}{{2^{R_s } I}}Y,}\\
&{\textrm{ }}\mathop {\max }\limits_{\scriptstyle k \in {\cal U} \hfill \atop
  \scriptstyle k \ne i \hfill} |h_{kb} |^2  < X
\end{split}
\right)}.\tag{C.1}\label{C.1}
\end{equation}
Considering that the fading coefficients $|h_{ie_j}|^2$ for $e_j \in {\cal E}$ are i.i.d. with the same mean of $\sigma^2_{ie}$, we readily obtain that the random variable $z$ is Gamma distributed with the mean of $N\sigma^2_{ie}$. Since $|{h_{ib}}{|^2}$ and $|{h_{ip}}{|^2}$ are independent and exponentially distributed with respective means of $\sigma^2_{ib}$ and $\sigma^2_{ip}$, we obtain the joint PDF of $(X,Y,Z)$ as
\begin{equation}
f (x,y,z) = \frac{{z^{N - 1} }}{{\Gamma (N)\sigma _{ib}^2 \sigma _{ip}^2 \sigma _{ie}^{2N} }}\exp ( - \frac{x}{{\sigma _{ib}^2 }} - \frac{y}{{\sigma _{ip}^2 }} - \frac{z}{{\sigma _{ie}^2 }}),\tag{C.2}\label{C.2}
\end{equation}
for $(x,y,z)>0$. Noting that $|{h_{kb}}{|^2}$ and $|{h_{ib}}{|^2}$ are independent exponential random variables and combining (C.1) and (C.2), we have
\begin{equation}
P_{out}^{sub}  = \sum\limits_{i = 1}^M {\iiint\limits_{\Theta} {{\prod\limits_{\scriptstyle k \in {\cal U} \hfill \atop
  \scriptstyle k \ne i \hfill} {\left( {1 - \exp ( - \frac{x}{{\sigma _{kb}^2 }})} \right)} f(x,y,z)dxdydz}}}
,\tag{C.3}\label{C.3}
\end{equation}
where $\Theta  =\left\{ {(x,y,z)|z > \frac{{N_e }}{{2^{R_s } N_b }}x - \frac{{(2^{R_s }  - 1)N_e }}{{2^{R_s } I}}y} \right\}$. Noting $(x,y,z)>0$, we may divide $\Theta$ into two mutually exclusively sets i.e. $\Theta _1$ and $\Theta _2$, where $\Theta _1$ and $\Theta _2$ are given by
\begin{equation}
\Theta _1  = \left\{ {(x,y,z) \left|{
\begin{split}
&N_b^{ - 1} Ix - (2^{R_s }  - 1)y > 0,\\
&z > \frac{{N_e }}{{2^{R_s } N_b }}x - \frac{{(2^{R_s }  - 1)N_e }}{{2^{R_s } I}}y
\end{split}
}\right. }\right\},\tag{C.4}\label{C.4}
\end{equation}
and
\begin{equation}
\Theta _2  = \left\{ {(x,y,z)|N_b^{ - 1} Ix - (2^{R_s }  - 1)y < 0,z > 0} \right\}.\tag{C.5}\label{C.5}
\end{equation}
By using $(x,y,z)>0$, $\Theta _1$ can be further divided into two mutually exclusively sets $\Theta _{11}$ and $\Theta _{12}$, which are described as
\begin{equation}
\Theta _{11}  = \left\{ {(x,y,z)\left|{
\begin{split}
&\frac{{Ix}}{{(2^{R_s }  - 1)N_b }} > y,\\
&y> \frac{{Ix}}{{(2^{R_s }  - 1)N_b }} - \frac{{2^{R_s } Iz}}{{(2^{R_s }  - 1)N_e }},\\
&\frac{{N_e }}{{2^{R_s } N_b }}x - z > 0
\end{split}
}\right.
} \right\},\tag{C.6}\label{C.6}
\end{equation}
and
\begin{equation}
\Theta _{12}  = \left\{ {(x,y,z)\left|{
\begin{split}
&\frac{{Ix}}{{(2^{R_s }  - 1)N_b }} > y > 0,\\
&\frac{{N_e }}{{2^{R_s } N_b }}x - z < 0
\end{split}
}\right.
} \right\}
.\tag{C.7}\label{C.7}
\end{equation}
Substituting $\Theta  = \Theta _{11}  \cup \Theta _{12}  \cup \Theta _2$ into (C.3) yields
\begin{equation}
P_{out}^{sub}  = \sum\limits_{i = 1}^M {(P_{out,I}  + P_{out,II}  + P_{out,III} )},
\tag{C.8}\label{C.8}
\end{equation}
where $P_{out,I}$, $P_{out,II}$ and $P_{out,III} $ are given by
\begin{equation}
P_{out,I}  = \sum\limits_{i = 1}^M {\iiint\limits_{\Theta_{11}} {{\prod\limits_{\scriptstyle k \in {\cal U} \hfill \atop
  \scriptstyle k \ne i \hfill} {\left( {1 - \exp ( - \frac{x}{{\sigma _{kb}^2 }})} \right)} f(x,y,z)dxdydz}}}
,\tag{C.9}\label{C.9}
\end{equation}
and
\begin{equation}
P_{out,II}  = \sum\limits_{i = 1}^M {\iiint\limits_{\Theta_{12}} {{\prod\limits_{\scriptstyle k \in {\cal U} \hfill \atop
  \scriptstyle k \ne i \hfill} {\left( {1 - \exp ( - \frac{x}{{\sigma _{kb}^2 }})} \right)} f(x,y,z)dxdydz}}}
,\tag{C.10}\label{C.10}
\end{equation}
and
\begin{equation}
P_{out,III}  = \sum\limits_{i = 1}^M {\iiint\limits_{\Theta_{2}} {{\prod\limits_{\scriptstyle k \in {\cal U} \hfill \atop
  \scriptstyle k \ne i \hfill} {\left( {1 - \exp ( - \frac{x}{{\sigma _{kb}^2 }})} \right)} f(x,y,z)dxdydz}}}
.\tag{C.11}\label{C.11}
\end{equation}
Substituting (C.2) and (C.6) into (C.9) yields
\begin{equation}
\begin{split}
P_{out,I}  =& \int_0^\infty  {\frac{{z^{N - 1} }}{{\Gamma (N)\sigma _{ie}^{2N} }}\exp ( - \frac{z}{{\sigma _{ie}^2 }})dz}\\
&\times\int_{\frac{{2^{R_s } N_b z}}{{N_e }}}^\infty  {\frac{1}{{\sigma _{ib}^2 }}\exp ( - \frac{x}{{\sigma _{ib}^2 }})\prod\limits_{\scriptstyle k \in {\cal U} \hfill \atop
  \scriptstyle k \ne i \hfill} {\left( {1 - \exp ( - \frac{x}{{\sigma _{kb}^2 }})} \right)} dx}   \\
&\quad\times \int_{\frac{{Ix}}{{(2^{R_s }  - 1)N_b }} - \frac{{2^{R_s } Iz}}{{(2^{R_s }  - 1)N_e }}}^{\frac{{Ix}}{{(2^{R_s }  - 1)N_b }}} {\frac{1}{{\sigma _{ip}^2 }}\exp ( - \frac{y}{{\sigma _{ip}^2 }})dy},  \\
\end{split}\tag{C.12}\label{C.12}
\end{equation}
where the term $\prod\limits_{\scriptstyle k \in {\cal U} \hfill \atop\scriptstyle k \ne i \hfill} {\left( {1 - \exp ( - \frac{x}{{\sigma _{kb}^2 }})} \right)}$ may be expanded to $ \sum\limits_{m = 0}^{2^{M - 1}  - 1} {( - 1)^{|{\cal U}_m |} \exp ( - \sum\limits_{k \in {\cal U}_m } {\frac{x}{{\sigma _{kb}^2 }}} )}$ by using the Binomial theorem, wherein ${\sum\limits_{k \in {\cal U}_m } {\frac{x}{{\sigma _{kb}^2 }}} }=0$ for ${\cal U}_m={\cal U}_0$ and ${\cal U}_0$ represents an empty set. Substituting this result into (C.12), we obtain (C.13) at the top of the following page.
\begin{figure*}
\begin{equation}
\begin{split}
P_{out,I}  = \frac{{( - 1)^{|{\cal U}_m |} (\frac{1}{{\sigma _{ie}^2 }} + \frac{{2^{R_s } N_b }}{{\sigma _{ib}^2 N_e }} + \sum\limits_{k \in {\cal U}_m } {\frac{{2^{R_s } N_b }}{{\sigma _{kb}^2 N_e }}} )^{ - N} }}{{\sigma _{ib}^2 \sigma _{ie}^{2N} [\sigma _{ib}^{ - 2}  + \sum\limits_{k \in {\cal U}_m } {\sigma _{kb}^{ - 2} }  + \frac{I(2^{R_s }  - 1)^{ - 1}} {\sigma _{ip}^{ 2} N_b} ]}} - \frac{{( - 1)^{|{\cal U}_m |} [\sigma _{ie}^{ - 2}  + \frac{{2^{R_s } N_b }}{{\sigma _{ib}^2 N_e }} + \sum\limits_{k \in {\cal U}_m } {\frac{{2^{R_s } N_b }}{{\sigma _{kb}^2 N_e }}}  + \frac{{I2^{R_s } }}{{(2^{R_s }  - 1)N_e \sigma _{ip}^2 }}]^{ - N} }}{{\sigma _{ib}^2 \sigma _{ie}^{2N} [\sigma _{ib}^{ - 2}  + \sum\limits_{k \in {\cal U}_m } {\sigma _{kb}^{ - 2} }  + I(2^{R_s }  - 1)^{ - 1} \sigma _{ip}^{ - 2} N_b^{ - 1} ]}} \\
\end{split}\tag{C.13}\label{C.13}
\end{equation}
\end{figure*}
Similarly, substituting (C.2) and (C.7) into (C.10) yields (C.14) at the top of the following page.
\begin{equation}
\begin{split}
P_{out,II}
&= \frac{{( - 1)^{|{\cal U}_m |} (\sigma _{ie}^{2N}  - C^{ - N} )}}{{\sigma _{ib}^2 \sigma _{ie}^{2N} (\sigma _{ib}^{ - 2}  + \sum\limits_{k \in {\cal U}_m } {\sigma _{kb}^{ - 2} } )}} \\
&\quad- \frac{{( - 1)^{|{\cal U}_m |} [\sigma _{ie}^{2N}  - (C  + \frac{{2^{R_s } I}}{{(2^{R_s }  - 1)\sigma _{ip}^2 N_e }})^{ - N} ]}}{{\sigma _{ib}^2 \sigma _{ie}^{2N} [B  + I(2^{R_s }  - 1)^{ - 1} \sigma _{ip}^{ - 2} N_b^{ - 1} ]}}, \\
\end{split}\tag{C.14}\label{C.14}
\end{equation}
where $B=\frac{1}{{\sigma _{ib}^2}} + \sum\limits_{k \in {{\cal{U}}_m}} {\frac{1}{{\sigma _{kb}^2}}}$ and $C=\frac{1}{{\sigma _{ie}^2 }} + \frac{{2^{R_s } N_b }}{{\sigma _{ib}^2 N_e }} + \sum\limits_{k \in {\cal U}_m } {\frac{{2^{R_s } N_b }}{{\sigma _{kb}^2 N_e }}} $. Additionally, substituting (C.2) and (C.5) into (C.11), we have
\begin{equation}
\begin{split}
 P_{out,III}
= \sum\limits_{m = 0}^{2^{M - 1}  - 1} {( - 1)^{|U_m |} \left( {\sigma^2_{ib}B + \frac{{I\sigma _{ib}^2 }}{{(2^{R_s }  - 1)\sigma _{ip}^2 N_b }}} \right)^{ - 1} },  \\
\end{split}\tag{C.15}\label{C.15}
\end{equation}
where $B=\frac{1}{{\sigma _{ib}^2}} + \sum\limits_{k \in {{\cal{U}}_m}} {\frac{1}{{\sigma _{kb}^2}}}$. This completes the derivation of (28).
}

\section{Proof of (55)}
Without loss of generality, let $z$ denote $\frac{x}{{\sigma _{kb}^2}}$, i.e., $ z = \frac{x}{{\sigma _{kb}^2}}$. Thus, the mean of random variable $z$ is obtained as
\begin{equation}
E(z) = \frac{1}{{\sigma _{kb}^2}}E(x),\tag{D.1}\label{D.1}
\end{equation}
which can be further computed from (56) as
\begin{equation}
\begin{split}
 E(z) &= \frac{1}{{\sigma _{kb}^2}}\int_0^\infty  {\frac{x}{{\sigma _{ib}^2}}\exp ( - \frac{x}{{\sigma _{ib}^2}} - \sum\limits_{{e_j} \in {{\cal {E}}_n}} {\frac{{{N_{{e_j}}}x}}{{\sigma _{i{e_j}}^2{2^{{R_s}}}{N_b}}}} )dx}  \\
&= \frac{1}{{\sigma _{kb}^2\sigma _{ib}^2}}{\left( {\frac{1}{{\sigma _{ib}^2}} + \sum\limits_{{e_j} \in {{\cal {E}}_n}} {\frac{{{N_{{e_j}}}}}{{\sigma _{i{e_j}}^2{2^{{R_s}}}{N_b}}}} } \right)^{ - 2}}.
\end{split}\tag{D.2}\label{D.2}
\end{equation}
Denoting $\sigma _{kb}^2 = {\theta _{kb}}\sigma _m^2$, $\sigma _{ib}^2 = {\theta _{ib}}\sigma _m^2$ and $\sigma _{i{e_j}}^2 = {\theta _{i{e_j}}}\sigma _e^2$, we can rewrite (D.2) as
\begin{equation}
E(z) = \frac{1}{{{\theta _{kb}}{\theta _{ib}}}}{\left( {\frac{1}{{{\theta _{ib}}{\lambda _{me}}}} + \sum\limits_{{e_j} \in {{\cal {E}}_n}} {\frac{{{N_{{e_j}}}}}{{{\theta _{i{e_j}}}{2^{{R_s}}}{N_b}}}} } \right)^{ - 2}} \cdot {(\frac{1}{{{\lambda _{me}}}})^2},\tag{D.3}\label{D.3}
\end{equation}
where ${\lambda _{me}} = \frac{{\sigma _m^2}}{{\sigma _e^2}}$. Meanwhile, we can also obtain the mean of $z^2$ as
\begin{equation}
\begin{split}
 E({z^2})
= \frac{1}{{\theta _{kb}^2{\theta _{ib}}}}{\left( {\frac{1}{{{\theta _{ib}}{\lambda _{me}}}} + \sum\limits_{{e_j} \in {{\cal {E}}_n}} {\frac{{{N_{{e_j}}}}}{{{\theta _{i{e_j}}}{2^{{R_s}}}{N_b}}}} } \right)^{ - 3}} \cdot {(\frac{1}{{{\lambda _{me}}}})^3}. \\
\end{split}\tag{D.4}\label{D.4}
\end{equation}
One can observe from (D.3) and (D.4) that for $\lambda_{me} \to \infty$, both $E(z)$ and $E(z^2)$ tend to zero. This implies that as $\lambda_{me} \to \infty$, random variable $z$ approaches zero with probability 1, yielding $z \buildrel 1 \over = 0$ for ${{\lambda _{me}} \to \infty }$, where $\buildrel 1 \over =$ denotes an equality with probability 1. Also, from the Maclaurin series expansion and the Cauchy's Mean-Value theorem, we obtain
\begin{equation}
1 - \exp ( - z) = z + \frac{{{z^2}}}{2}\exp ( - \theta z ),\label{D.5}\tag{D.5}
\end{equation}
where $0<\theta<1$. From (D.5), we have
\begin{equation}
\mathop {\lim }\limits_{{\lambda _{me}} \to \infty } 1 - \exp ( - z) = z + \mathop {\lim }\limits_{{\lambda _{me}} \to \infty } \frac{{{z^2}}}{2}\exp ( - \theta z). \label{D.6}\tag{D.6}
\end{equation}
Similarly to (D.3) and (D.4), we can easily prove that for ${{\lambda _{me}} \to \infty }$, the mean and variance of $z^2$ are high-order infinitesimals as compared with the mean and variance of $z$. Meanwhile, due to $0<\theta<1$ and $z>0$, we have $0< \exp ( - \theta z)<1$. Thus, term $\frac{{{z^2}}}{2}\exp ( - \theta z )$ is high-order infinitesimal compared to $z$, as $\lambda_{me} \to \infty$. Ignoring the high-order infinitesimal in (D.6) yields
\begin{equation}
\mathop {\lim }\limits_{{\lambda _{me}} \to \infty } 1 - \exp ( - z) \buildrel 1 \over= z.\label{D.7}\tag{D.7}
\end{equation}
Substituting $ z = \frac{x}{{\sigma _{kb}^2}}$ into (D.7) yields
\begin{equation}
1 - \exp ( - \frac{x}{{\sigma _{kb}^2}}) \mathop  = \limits^1 \frac{x}{{\sigma _{kb}^2}},\tag{D.8}\label{D.8}
\end{equation}
for $\lambda_{me} \to \infty$, which is (55).

\begin{IEEEbiography}[{\includegraphics[width=1in,height=1.25in]{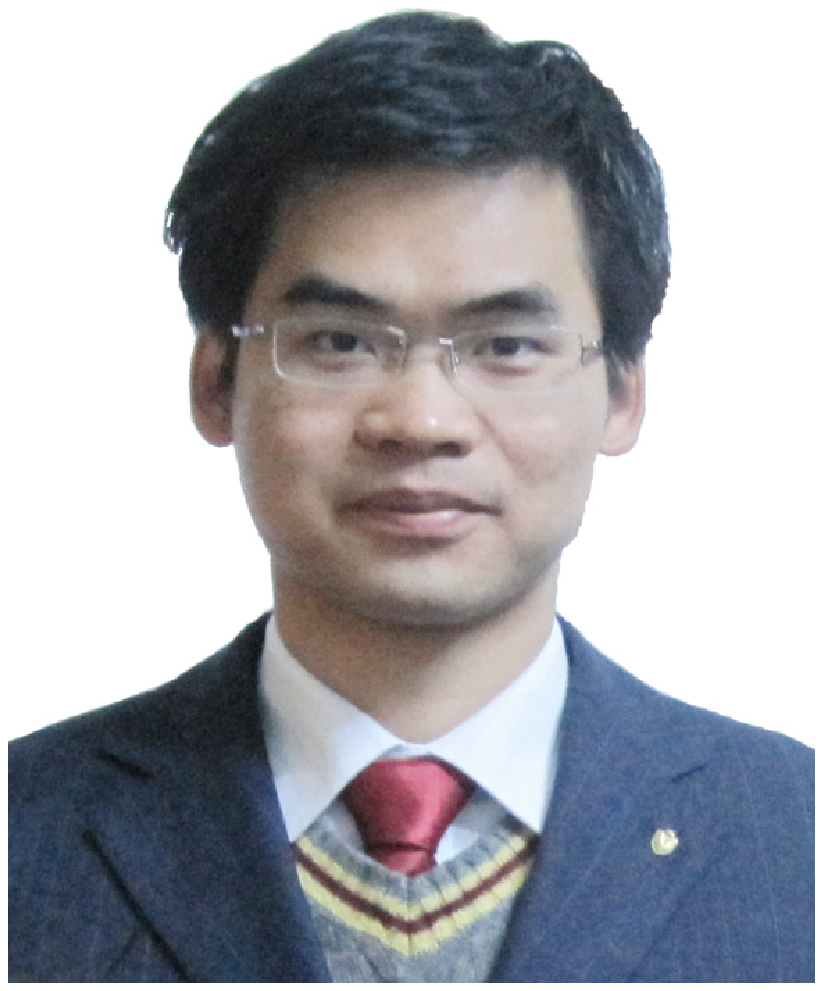}}]{Yulong Zou} (S'07-M'12-SM'13) is a Full Professor at the Nanjing University of Posts and Telecommunications (NUPT), Nanjing, China. He received the B.Eng. degree in Information Engineering from NUPT, Nanjing, China, in July 2006, the first Ph.D. degree in Electrical Engineering from the Stevens Institute of Technology, New Jersey, United States, in May 2012, and the second Ph.D. degree in Signal and Information Processing from NUPT, Nanjing, China, in July 2012. His research interests span a wide range of topics in wireless communications and signal processing, including the cooperative communications, cognitive radio, wireless security, and energy-efficient communications.

Dr. Zou is currently serving as an editor for the IEEE Communications Surveys \& Tutorials, IEEE Communications Letters, EURASIP Journal on Advances in Signal Processing, and KSII Transactions on Internet and Information Systems. He has also served as the lead guest editor for a special issue on ``Security Challenges and Issues in Cognitive Radio Networks" in the EURASIP Journal on Advances in Signal Processing. In addition, he has acted as symposium chairs, session chairs, and TPC members for a number of IEEE sponsored conferences, including the IEEE Wireless Communications and Networking Conference (WCNC), IEEE Global Communications Conference (GLOBECOM), IEEE International Conference on Communications (ICC), IEEE Vehicular Technology Conference (VTC), International Conference on Communications in China (ICCC), and so on.
\end{IEEEbiography}

\begin{IEEEbiography}{Xuelong Li} (M'02-SM'07-F'12) is a Full Professor with the Center for OPTical IMagery Analysis and Learning (OPTIMAL), State Key Laboratory of Transient Optics and Photonics, Xi'an Institute of Optics and Precision Mechanics, Chinese Academy of Sciences, Xi'an 710119, Shaanxi, P. R. China.
\end{IEEEbiography}

\begin{IEEEbiography}[{\includegraphics[width=1in,height=1.25in]{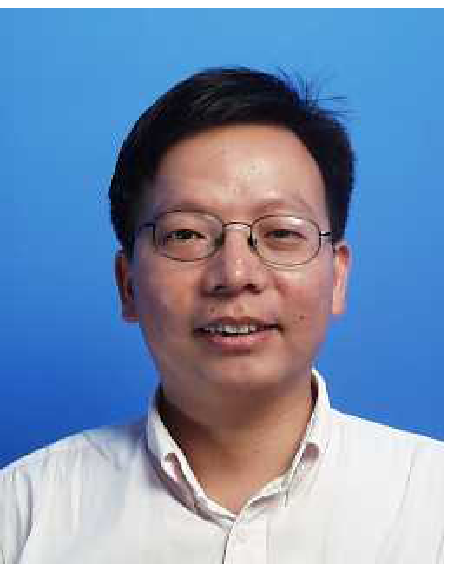}}]
{Ying-Chang Liang} (SM'00-F'11) is a Principal Scientist at the Institute for Infocomm Research (I2R), Agency for Science, Technology and Research (A*STAR), Singapore. He was a visiting scholar with the Department of Electrical Engineering, Stanford University, from December 2002 to December 2003, and taught graduate courses in National University of Singapore from 2004-2009. His research interest includes cognitive radio networks, dynamic spectrum access, reconfigurable signal processing for broadband communications, information theory and statistical signal processing.

Dr. Liang was elected a Fellow of the IEEE in 2011 for contributions to cognitive radio communications, and has received five Best Paper Awards, including IEEE ComSoc APB outstanding paper award in 2012, and EURASIP Journal of Wireless Communications and Networking best paper award in 2010. He also received the Institute of Engineers Singapore (IES)'s Prestigious Engineering Achievement Award in 2007, and the IEEE Standards Association's Certificate of Appreciation Award in 2011, for contributions to the development of IEEE 802.22, the first worldwide standard based on cognitive radio technology.

Dr. Liang currently serves as Editor-in-Chief of the IEEE JOURNAL ON SELECTED AREAS IN COMMUNICATIONS-Cognitive Radio Series, and is on the editorial board of the IEEE Signal Processing Magazine. He was an Associate Editor of the IEEE TRANSACTIONS ON WIRELESS COMMUNICATIONS and the IEEE TRANSACTIONS ON VEHICULAR TECHNOLOGY, and served as a Guest Editor of five special issues on emerging topics published in IEEE, EURASIP and Elsevier journals in the past years. He is a Distinguished Lecturer of the IEEE Communications Society and IEEE Vehicular Technology Society, and is a member of the Board of Governors of the IEEE Asia-Pacific Wireless Communications Symposium. He served as technical program committee (TPC) Co-Chair of 2010 IEEE Symposium on New Frontiers in Dynamic Spectrum Access Networks (DySPAN'10), General Co-Chair of 2010 IEEE International Conference on Communications Systems (ICCS'10), and Symposium Chair of 2012 IEEE International Conference on Communications (ICC'12).
\end{IEEEbiography}

\end{document}